# TRACKING NETWORK DYNAMICS: A SURVEY OF DISTANCES AND SIMILARITY METRICS

By Claire Donnat and Susan Holmes

*Department of Statistics, Stanford University*†

From longitudinal biomedical studies to social networks, graphs have emerged as a powerful framework for describing evolving interactions between agents in complex systems. In such studies, after pre-processing, the data can be represented by a set of graphs, each graph represents a system's state at a different point in time or space. The analysis of the system's dynamics depends on the selection of the appropriate analytical tools. In particular, after specifying properties characterizing similarities between states, a critical step lies in the choice of a distance between graphs capable of reflecting such similarities.

While the literature offers a number of distances that one could a priori choose from, their properties have been little investigated and no guidelines regarding the choice of such a distance have yet been provided. In particular, most graph distances consider that the nodes are exchangeable and do not take into account node identities. Accounting for the alignment of the graphs enables us to enhance these distances' sensitivity to perturbations in the network and detect important changes in graph dynamics. Thus the selection of an adequate metric is a decisive – yet delicate – practical matter.

In the spirit of Goldenberg, Zheng and Fienberg's seminal 2009 review [21], the purpose of this article is to provide an overview of commonly-used graph distances and an explicit characterization of the structural changes that they are best able to capture. To see how this translates in real-life situations, we use as a guiding thread to our discussion the application of these distances to the analysis of both a longitudinal microbiome dataset and a brain fMRI study. We show examples of using permutation tests to detect the effect of covariates on the graphs' variability. Finally synthetic examples provide intuition as to the qualities and drawbacks of the different distances. Above all, we provide some guidance for choosing one distance over another in certain types of applications.

Finally, extending the scope of our analysis from temporal to spatial dynamics, we show an application of these different distances to a network created from worldwide recipes.

*Supported by NIH AI112401 and NSF DMS 1501767. The fMRI example was funded by a grant from the National Institute on Drug Abuse (R03DA024775) to Clare Kelly. This article was written while SH was a fellow at CASBS, whose support is gratefully acknowledged.

CONTENTS

## 1. Introduction and Motivation.

**Motivation.** From social sciences to biology, scientific communities across a wide number of disciplines have become increasingly interested in the study of networks – that is, graphs in which each entity or data point is assigned to a node, and existing interactions or similarities between entities are modeled by edges. If graphs provide a versatile framework for encapsulating structural information in datasets, they also come as an indispensable paradigm in a number of applications where the study of each individual node represented by Euclidean features is either irrelevant or intractable. Brain connectome data represent brain activity by modeling neurons' activation patterns from a network perspective – rather than by recording each individual neuron's activity. The focus of such studies is the system as a whole rather than at an atomic level. Similarly, in microbial ecology, communities of bacteria can be represented by a co-occurrence graphs where the edges are a (carefully-selected) function of bacterias' co-abundances. The representation of biological samples as graphs provides a richer, more informative framework than the bacterial counts themselves. Recent studies have associated "significant" bacterial communities to various medical conditions, such as obesity [54, 53, 25] or preterm birth [14].

1.1. *Application: microbiome and fMRI data .*
We are going to use two examples that illustrate all the different distances throughout for our discussion.
**The 2011 Relman antibiotics dataset.** This longitudinal microbiome study consists of a set of 162 bacterial samples taken from the gut of three distinct subjects (D, E, and F) at different points in time. The subjects were given two courses of antibiotics over ten months, yielding seven distinct treatment phases (pre-treatment, first antibiotic course, week after stopping treatment 1, interim, second course of antibiotics, week after stopping treatment 2, and post-treatment phase). The goal of the study was to assess the antibiotics' effects on microbial communities. While this dataset has been already analyzed in the literature [13, 19], we here propose to tackle it from a new network perspective. This analysis aims to provide complimentary information to the previous by allowing the analysis of higher-order interactions between bacteria: can we characterize prevalent communities of bacteria for each treatment phase? How do these communities react to the different drugs? The analysis of co-occurrence networks in microbiome samples is becoming increasingly popular [4, 20, 39, 44, 56]. A critical step is the transformation of the raw bacterial counts into a graph capturing such interactions. While a plethora of methods have been suggested for inferring networks from the abundance matrix (see Appendix A for more references), our analysis is done on bacteria "co-occurrence" networks. For each subject at a given treatment phase, we define a graph in which each node corresponds to a specific species, and edges $\mathcal{E} = \{(i, j)\}$ capture pairwise "affinities" between bacteria $i$ and $j$. Intuitively, these affinities capture symbiotic mechanisms between bacteria: do they thrive simultaneously together – or, on the contrary – does one bacteria tend to smother the development of another? Because of the particular nature of the data (zero-inflated negative binomial), we use a thresholded-Kendall correlation as a measure of the tendency of two bacteria to thrive (or wither) together. Indeed, the Kendall correlation is a ranked-based correlation and is as such less biased by the over-representation of zeros in the data. This allows us to obtain a set of twenty-one different graphs—one for each of the three subjects during each of the treatment phases— on the 2,582 nodes representing the different species. More details regarding the explicit construction of these graphs are given in Appendix A.

**Resting-State fMRI data.** We are also going to compare brain connectomes, which have radically different properties to the microbiome. The dataset consists in the resting-states fMRI data of 29 cocaine-dependent patients, and was published as part as a study of the effect of cocaine addiction on functional and structural connectivity[33][1]. In their 2011 article, the authors of the study show the existence of a statistically-significant reduction in the interhemispheric connectivity between cocaine-users and controls, highlighting the existence of an effect of substance-abuse on functional connectivity. In a similar spirit, we apply different network distances to this dataset in order to assess if the number of years under cocaine-dependence correlates with differences between the different connectomes. Each patient's fMRI raw data has been preprocessed through the FSL standard pipeline (in particular, the fMRI have been registered to a template brain, and each voxel's time series has been scrubbed to account for small head movements, and the "global signal" has been regressed out of each time series. More details can be found in Appendix B). This filtering procedure yields a total of 116 nodes, with values over 140 time points. To create a (weighted) graph from these filtered time series, we then compute the Pearson correlation and keep only correlations above a given level. Our approach is akin to [50] who select this threshold by controlling the number of edges in the graphs: the threshold used here is the mean (across subject) of each correlation matrix's $97^{th}$ quantile. In average, the graphs that we recover have around 3% sparsity.

Given these sets of graphs, the crux of the analysis lies in the choice of a distance capable of identifying similar "graph-states", tailored to the data at hand. In the microbiome example, while the thousands of taxa that constitute the human microbiome allow for a rich variety of potentially different microbial communities, these communities usually involve a very small proportion of the taxa, and the associated abundance matrices are typically very sparse (here, 17% of the observations are non-zero.) In the brain connectome setting, while the raw fMRI data is "better behaved" and follows a normal distribution more closely, it is also usually characterized by a high amount of noise: preprocessing steps, such as the template alignment and Gaussian smoothing performed to realign the brains and account for small head movements during the data retrieval, are known to blur even further the signal-to-noise ratio.

[1] Data publicly available at the following link: http://fcon_1000.projects.nitrc.org/indi/ACPI/html/acpi_nyu_1.html

### 1.2. *Problem statement and notation.*

**Problem statement.** We now assume that the raw data has been transformed into a set of different graphs, which are considered the essential summaries of the data. Hereon, our goal is to assess the differences between these summaries and we treat them as our input. In this perspective, the definition of a distance between aligned graphs –that is, graphs defined on the same set of identified nodes– takes on a significant importance. The classical problem of assessing the distance between two unlabeled graphs has been well studied[9, 36, 45, 58], our focus is different. Indeed, since in our examples the nodes have been endowed with a particular identity, there is no need to consider permutation-invariant distances. On the contrary, one might even wish to leverage the information contained in the nodes' labeling to define a distance sensitive to the intensity of changes at key-node locations. However, while the literature provides us with a number of "off-the-shelf" graph distances – any of which being, in principle, suited for the task–, these distances exhibit in fact distinct properties and capture different types of structural changes. For example, assessing the amount of overall network changes typically calls for a different distance than if changes are weighted by their degree of importance or impact on the structure (edges addition in sparse areas of the graph yield less redundancies and can be considered as more crucial than in densely connected areas for instance). We hope to provide some elements that guide the selection of the most appropriate distance given the analyst's desiderata.

This review investigates the properties of some of the many different distances and similarities between graphs. We will include pseudo-distances – that is, metrics such that $d(G_1, G_2) = 0$ does not necessarily imply that graphs $G_1$ and $G_2$ are strictly the same, but rather that they share identical

"key" characteristics. Starting with structural and spectral distances –which constitute the main bulk of metrics proposed in the literature–, we highlight the dynamics and types of structural changes that these two main categories are best able to capture. An improved understanding of these metrics' properties suggests new similarities tailored to specific scenarios and address some of the mainstream distances' shortcomings. We introduce a new set of graph similarities based on spectral heat kernels, and argue that these similarities are optimal in that they combine both local and global structural information. In each case, the performance of the different distances is assessed on both the microbiome and fMRI datasets, as well as on synthetic, controlled experiments. Finally, we extend our illustration by applying them to the analysis of a "recipe" networks.

**Notation.** Throughout this review, we write $G = (\mathcal{V}, \mathcal{E})$ the graph with vertices $\mathcal{V}$ and edges $\mathcal{E}$. We denote as $N = |\mathcal{V}|$ the number of nodes, $|\mathcal{E}|$ the number of edges, and we write $i \sim j$ if nodes $i$ and $j$ are neighbors. Our framework considers undirected binary graphs, with no self loops (which we extend to the study of weighted graphs in our applications). $A$ refers to the adjacency matrix of the graph, and $D$ to its degree matrix:

$$A_{ij} = \begin{cases} 1 \text{ if } i \sim j \\ 0 \text{ otherwise} \end{cases} \qquad \text{and} \qquad D = \text{Diag}(d_i)_{i=1\cdots N} \quad \text{s.t.} \quad d_i = \sum_{j=1}^{N} A_{ij}$$

In our case of undirected graphs, the matrix $A$ is symmetric: $A^T = A$. Following standard graph theory [6],the Laplacian of the graph is the matrix defined as: $L = D - A$. The Laplacian is symmetric, and we consistently write its (real-valued) eigenvalue decomposition as : $L = U\Lambda U^T$, where $U$ is a unitary matrix, and $\Lambda = \text{Diag}(\lambda_i)$ is the diagonal matrix of the eigenvalues: $0 = \lambda_0 \leq \lambda_1 \leq \cdots \leq \lambda_{N-1}$.

## 2. Quantifying local changes via structural distances.

Distances between graphs usually fall in one of two general categories, often considered as mutually exclusive: structural vs. spectral distances. The first one captures local changes, whereas the second one assesses the smoothness of the evolution of the overall graph structure by tracking changes in the eigenvalues of the graph Laplacian or its adjacency matrix. We begin our review by analyzing properties of these two types of distances.

### 2.1. *The Hamming distance.*

**Definition.** The Hamming distance – a special instance of the broader class of Graph-Edit distances – measures the number of edge deletions and insertions necessary to transform one graph into another. More formally, let $G$ and $\tilde{G}$ be two graphs on $N$ nodes, as well as $A$ and $\tilde{A}$ their corresponding adjacency matrices, the (normalized) Hamming distance is defined as :

$$d_H(G, \tilde{G}) = \sum_{i,j} \frac{|A_{ij} - \tilde{A}_{ij}|}{N(N-1)} = \frac{1}{N(N-1)} ||A - \tilde{A}||_{1,1} \tag{2.1}$$

This defines a metric between graphs, since it is a scaled version of the $L_{1,1}$ norm between the adjacency matrices $A$ and $\tilde{A}$. As such, Eq. 2.1 defines a distance bounded between 0 and 1 over all graphs of size $N$.

**Application.** Figure 1 illustrates the results of the analysis of the microbiome study using the Hamming distance on both the bacterial (top row), as well as the fMRI data (bottom row-right).

**General analysis framework.** We briefly outline here the framework that we use throughout the paper to analyze these graphs through each dissimilarity. For each dataset (microbiome and fMRI) and each distance, we store the pairwise dissimilarities between graphs in a $n$-by-$n$ dimensional matrix $H$, where $H_{ij} = d(G_i, G_j)$ and $n$ is the number of graphs ($n = 21$ in the microbiome example and $n = 29$ in the fMRI dataset). We use this matrix $H$ to analyze the relationship between H and a set of different factors. In particular, in the microbiome example, each distance is used for the analysis of: (a) the graphs' variability from time frame to time frame, illustrated by plots of distances between consecutive graphs (Figures 1B,1E) and (b) similarities across subjects or across treatment phases, illustrated by both heatmaps of the pairwise distances $H$ between graphs (Figures 1A,1E) and their low dimensional projections (multidimensional scaling MDS). Figures 1B,1D show such two dimensional projections. In the fMRI example, the analysis focuses on the relationship between the dissimilarities and the number of years under dependency

**Results: microbiome data.** In the antibiotic study (top row of Figure 1), the Hamming distance is computed between the graphs built between taxa at different stages of the time course. It shows the existence of similar dynamics across subjects, as highlighted by the closely matching shapes of the curves (Figure 1C) representing the evolution of the distances between consecutive graphs. The MDS projection (Figure 1B) on the first components highlights the existence of a "treatment gradient": interim phases –located in the bottom right corner of the figure– are closer to the pre-treatment samples and far from the treatment phases (violet and black points at center-left of the figure),consistent with biological interpretation of the treatment effects. While the Hamming distance does not detect stronger similarities between samples belonging to the same individual (no darker blue blocks along the diagonal of the heat map in Figure 1A), it is able to identify similar dynamic regimes across subjects – as highlighted by the clustered MDS projections of points corresponding to the same treatment phase. In order to quantify this effect, we run a Friedman-Rafsky test on the induced $k$-nearest-neighbor graph: for a given value of $k$, we compute the $k$-nearest-neighbor "metagraph" (or graph of graphs) induced by the pairwise-dissimilarity matrix $H$. This provides a useful way of extracting information from $H$ by representing it as a graph where each node is itself a graph, and edges reflect the $k$-strongest similarities between graphs. Henceforward, we refer to this induced $k$-nearest neighbor graph of graphs as the k-nn metagraph. Having constructed the k-nn metagraph, we compute the number of its edges which connect graphs of the same class (i.e, in the microbiome dataset, treatment stage or subject). We then permute the labels to generate 50,000 graphs with the same topology, but where the edges randomly connect nodes independently of their class. We compare the original value to this synthetic null permutation distribution to get the associated p-value: this assesses the compatibility of the distance on a given set of labels: if the distance clusters together graphs belonging to the same category, then the p-value should be significantly small. The p-values are reported alongside the plots in Figure 1(F), where we have conducted this experiment with $k = 1$ (the nearest-neighbor graph is thus simply the minimum spanning tree). In this case, interestingly, this test fails to report any statistically significant association between the edges in the minimum spanning tree and neither treatment stages nor the subject labels. We note that increasing the number of neighbors considered ($k = 2, 3..$) does not uncover any meaningful statistical associations between the induced topological ordering of the graphs and any of the node labels.

**Results: fMRI data.** When applied to the resting state-fMRI data, the Hamming distance does not detect any clusters of closely related graphs (as shown by the uniform cluster map in Figure 1E and the uniform tSNE projections in Figure 1D). We also adapt our previous Friedman-Rafksy test to handle continuous labels instead of discrete classes. This enables use to test the association of the k-nearest neighbor graph between patients and the amount of time that they have spent under cocaine dependency. The test statistic is now defined as the sum of the differences between labels for all the edges in the graph. In this setting, a small difference would indicate that brain networks are more similar to other networks with similar "time under dependence". As shown in both the figures and the plots, the Hamming distance does not detect any significant relationship between the relative distance of the graphs and their labels. We also run an analysis-of-variance type test: we split the graphs into two classes (with roughly the same number of subjects): patients with less than 5 years under cocaine dependency and patients with more than five years. We then compute the ratio $\Delta = \frac{\bar{D}_{12}}{\frac{n_1}{n_1+n_2}\bar{D}_{11}+\frac{n_2}{n_1+n_2}\bar{D}_{22}}$ where $\bar{D}_{ij}$ denotes the average distance between subjects of class $i$ and $j$: under the null, this ratio should be centered around 1. We assess the significance of this ratio via a permutation test, which here yields a p-value of 0.62: in this case, the Hamming distance does not detect any significant difference between the graphs in the two classes.

**Discussion.** With a cost complexity of the order of $O(N^2)$, the Hamming distance provides a straightforward way of comparing sequences of aligned graphs that only takes into account the number of shared edges. It thus comes as no surprise that this distance has been a long-time favorite in various graph comparison problems. Graph embedding techniques – which provide a vector-valued representation for each graph that captures its geometric properties– are a case in point: in [41], the authors define similarities between subgraphs through their graph-edit distance. Similarly, in [17], the authors introduce the notion of a "median graph" as the minimizer of the sum of pairwise graph-edit distances.

While the Hamming distance is a perfectly valid first candidate graph distance for any type of analysis, it is worth emphasizing that it only reveals some restricted aspect of network similarities.

The first trait to highlight is its uniform treatment of all changes in the graph structure: all additions and deletions are assumed to have similar importance. Changes in the network's core are treated equivalently to changes in the periphery. We will analyze the consequences and limitations of this assumption in section 2.1.3. A second trait is Hamming's sensitivity to the density of the graphs. This yields a limited capacity to recognize similar dynamical processes across graphs with varying sparsity . As an example of the first point, let us consider a dynamic regime in which, at every time point, each edge is randomly flipped independently of the others: it either stays in the graph or disappears with probability $p$. The total number of disappearances follows a binomial distribution with mean $p|\mathcal{E}|$. For an identical perturbation mechanism, dense graphs are thus placed at higher distances to each other – and are thus considered as more unstable – than sparse graphs. The Hamming distance is unable to recognize that these graphs share in fact the same level of relative variability, which can hinder some aspects of the analysis. Indeed, the random deletion process at hand can be thought of as blurring noise applied to a true underlying graph structure, and is a typical representation our inability to observe all interactions between nodes in a complex system. In this case, it seems more natural to specify the inherent variability of the data in terms of "noise level" rather than "noise quantity", and our analysis should thus recognize similar noise levels independently of the graphs' original sparsity. Similarly, the Hamming distance tends to place nested graphs at a smaller distance to each other than other metrics. Indeed, suppose that graph $\tilde{G}$ comprises 50% of the edges of graph $G$. The Hamming distance between the two graphs is then simply $d(G, \tilde{G}) = \frac{0.5||A||}{N(N-1)}$, and does not correct for the size of the initial graph. We could nonetheless argue that this distance should be big (or at least close to 0.5), since the structure of the system is radically modified. Our microbiome study is a case in point (Fig. 1B): the variety of the microbiota involved in the interim phase jumps to almost twice its corresponding value in any of the antibiotics phases (the number of bacteria increases from around 210 bacteria to 420). The distances between the interim phase and the other phases are subsequently smaller than for any of the other phases, with a number of shared taxa.

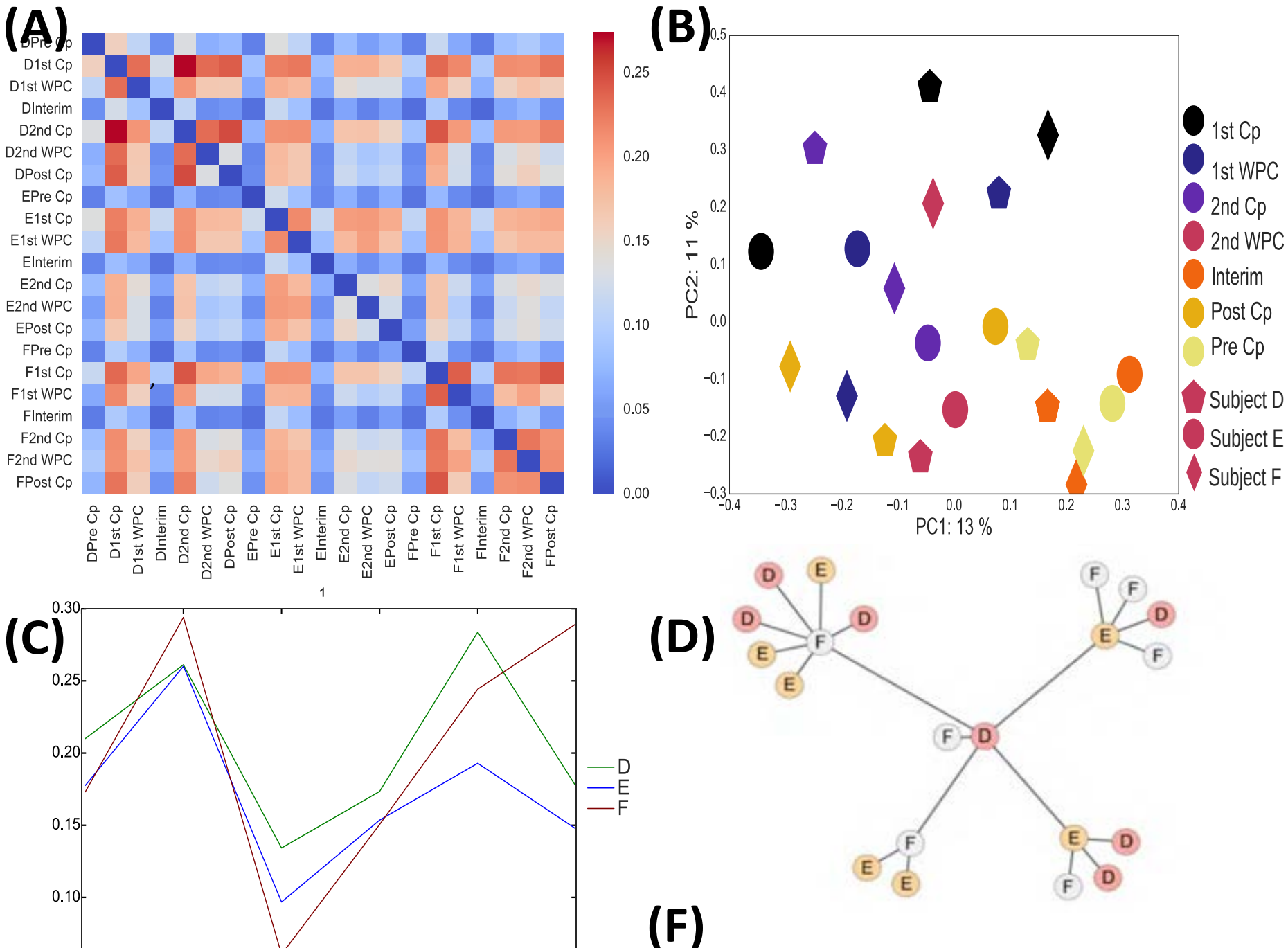

| Dataset and Labels | FR p-values |
|---|---|
| **Relman- Microbiome** (3 classes, one per subject) | 1.000 |
| **Relman- Microbiome** (7 classes, one per state) | 0.636 |
| **Kelly- fMRI dataset** (continuous labels, years of dependency) | 0.180 |

Table 1: Results of the Friedman-Rafsky test for the Minimum Spanning Tree induced by the Hamming dist.

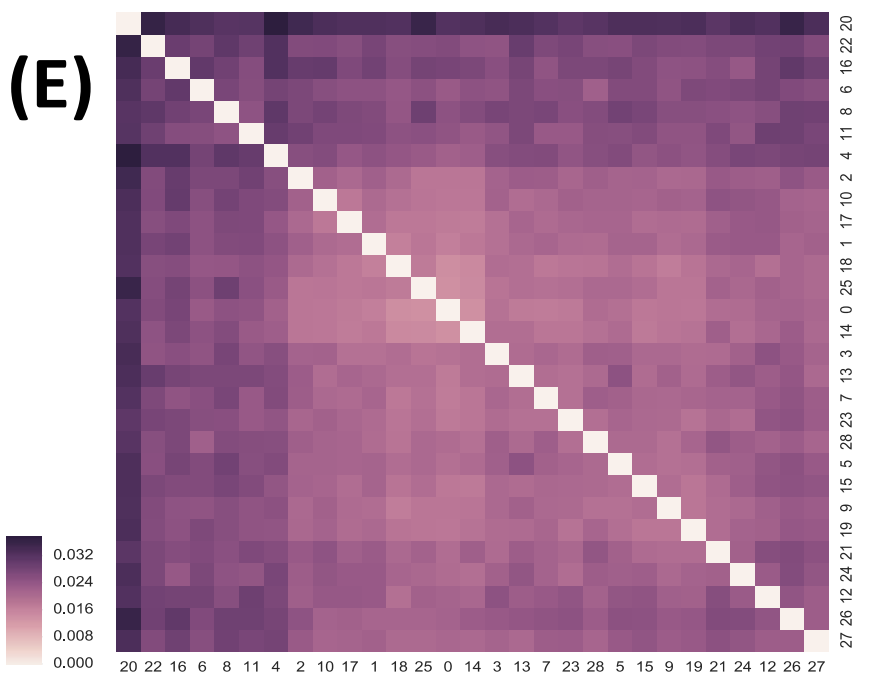

(G)

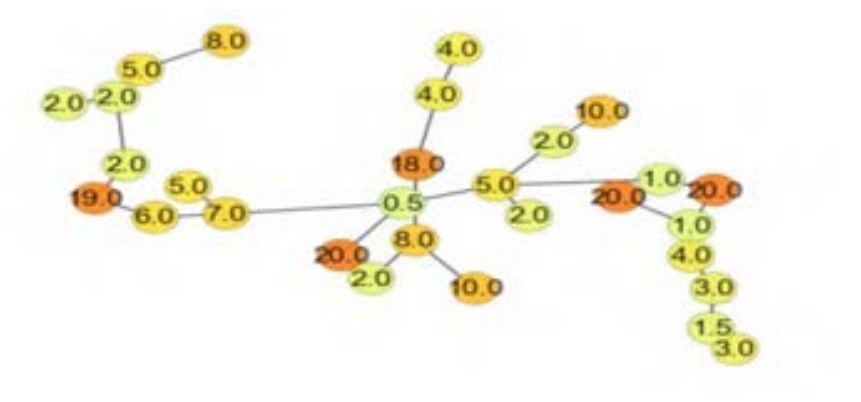

Fig 1: Hamming distance between bacterial graphs (top rows), and brain graphs (bottom row). Heatmap of the Hamming distances between Kendall-correlation-based bacterial graphs **(A)** and MDS projections on the first two principal components. **(B)**.Colors denote treatment phases, and shapes represent different subjects . Plots of the consecutive distances between bacterial graphs **(C)**. Minimum Spanning tree between bacterial graphs induced by the Hamming distance **(G)**. Friedman-Rafsky test for significance for the different datasets **(F)**. Clustermap of the fMRI graphs **(E)**. Minimum spanning tree between brain connectomes induced by the Hamming distance **(G)**.

The Hamming distance is thus a measure of the *amount of change* between two graphs. While this might be adequate for characterizing the evolution of a given system through time, it is nonetheless unfit for finding similarities in broader settings. Tasks such as comparing graph dynamics in the presence of different degree densities or recognizing instances of the same network family (Erdös-Rényi random graphs, preferential attachement graphs, etc.) indubitably require other metrics.

### 2.2. *The Jaccard distance.*

**Definition.** A potential solution to the aforementioned density-effect problem consists in using the Jaccard distance [40], which includes a normalization with respect to the volume of the union graph:

(2.2)
$$d_{\text{Jaccard}}(G, \tilde{G}) = \frac{|G \cup \tilde{G}| - |G \cap \tilde{G}|}{|G \cup \tilde{G}|} = \frac{\sum_{i,j} |A_{ij} - \tilde{A}_{ij}|}{\sum_{i,j} \max(A_{i,j}, \tilde{A}_{ij})}| = \frac{||A - \tilde{A}||_{1,1}}{||A + \tilde{A}||_{*}}$$

where $||\cdot||_*$ denotes the nuclear norm of a matrix.

Eq. 2.2 is known to define a proper distance between the graphs. A straightforward way to see this is to use the Steinhaus Transform, which states that for $(X, d)$ a metric and $c$ a fixed point, the transformation $\delta(x,y) = \frac{2d(x,y)}{d(x,c)+d(y,c)+d(x,y)}$ produces a metric. We apply here this transformation with $d$ the Hamming distance and $c$ the empty graph:

$$\delta(G,\tilde{G}) = \frac{2||A-\tilde{A}||_{1,1}}{||A||_{1,1}+||\tilde{A}||_{1,1}+||A-\tilde{A}||_{1,1}} = \frac{2(|G\cup\tilde{G}|-|G\cap\tilde{G}|)}{2|G\cup\tilde{G}|} \quad (*)$$
$$= d_{\text{Jaccard}}(G,\tilde{G}).$$

In particular, taking for instance $G = G_t$ and $\tilde{G} = G_{t+1}$ the graphs associated to the state of a system at two consecutive time points $t$ and $t+1$ (with $\mathcal{E}_{G_t}$ and $\mathcal{E}_{G_{t+1}}$ their respective set of undirected edges) and rewriting the left hand side of (*), we have:

$$\begin{aligned} d_{\text{Jaccard}}(G_t,G_{t+1}) &= \frac{d_{\text{Hamming}}(G_t,G_{t+1})}{\frac{|\mathcal{E}_{G_t}|+|\mathcal{E}_{G_{t+1}}|}{2N(N-1)}+\frac{1}{2}d_{\text{Hamming}}(G_t,G_{t+1})} \\ \Longrightarrow d_{\text{Jaccard}}(G_t,G_{t+1}) &= \frac{\frac{d_{\text{Hamming}}(G_t,G_{t+1})}{\bar{S}}}{1+\frac{d_{\text{Hamming}}(G_t,G_{t+1})}{2S}} \end{aligned} \tag{2.3}$$

with $\bar{S} = \frac{|\mathcal{E}_{G_t}|+|\mathcal{E}_{G_{t+1}}|}{2N(N-1)}$ is the average sparsity of the two graphs.

**Application.** Figure 2 shows the result of the analysis carried out using the Jaccard distance. Since the edges in our graphs have been assigned different weights according to the intensity of the interaction between bacteria, we have used the version of the Jaccard distance extended to the weighted graph setting, defined as:

$$d_{\text{Jaccard}}(G,\tilde{G}) = 1 - \frac{\sum_{i,j}\min(A_{ij},\tilde{A}_{ij})}{\sum_{i,j}\max(A_{ij},\tilde{A}_{ij})}.$$

This analysis yields somehow different results to the Hamming distance (Figures 2A,2D, 2C,2D). We note that the treatment phases express more variability and are far from most on the other samples. The Friedman-Rafsky test for the microbiome data highlighted a significant dependence of the 3-nn metagraph on the subject: with a p-value of 0.0002, this test shows that bacterial graphs corresponding to the same patient are significantly closer than under the random null model. This effect is further confirmed by running a analysis-of-variance type test and computing the statistics $\Delta = \frac{1}{3}\frac{\sum_{i\in\{D,E,F\}}\bar{D}_{i,i^c}}{\sum_{i\in\{D,E,F\}}\frac{n_i}{n_{tot}}\bar{D}_{i,i}}$ where $\bar{D}_{i,i^c}$ denotes the average distance between graphs in class $i$ and graphs in any other class. Under the null, this statistic

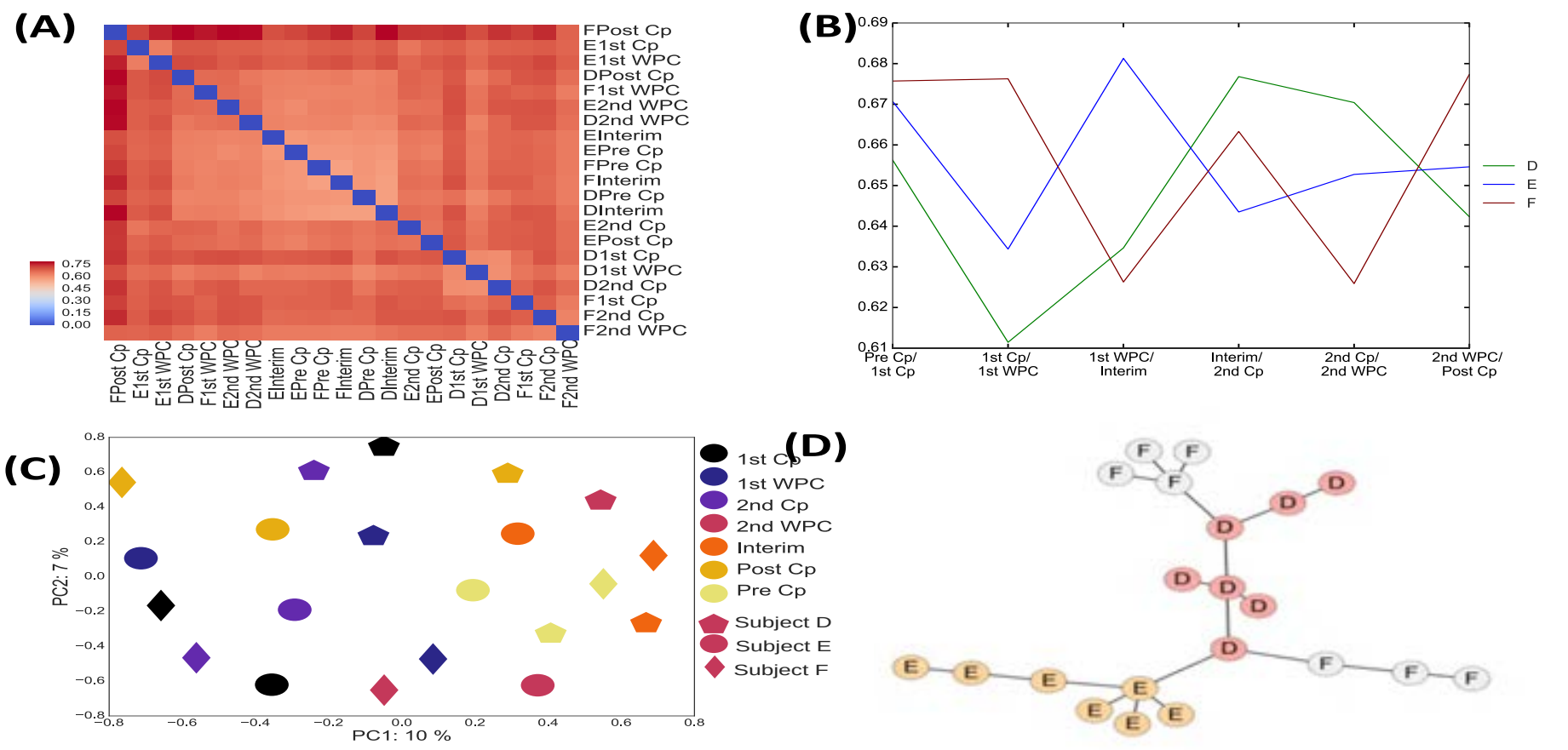

.

Fig 2: Application of the Jaccard distance to the microbiome study. Cluster-map of the Jaccard distances between Kendall-correlation-based bacterial graphs **(A)** . Plots of the consecutive distances between bacterial graphs **(B)**. MDS projection of the bacterial **(C)** graphs on the first two principal axes. Colors denote treatment phases, and shapes represent different subjects. Minimum spanning tree between bacterial graphs **(D)**

is centered at 1, and we evaluate its significance through a permutation test. This yields a p-value of 0.0018, highlighting the existence of a significant difference between graphs grouped according to their associated subject Id. This microbiome example is thus a case where the Jaccard distance is a better fit for our analysis: whereas the Hamming fails to uncover any real similarity between bacterial graphs corresponding to the same subject, the Jaccard distance does capture the existence of greater similarities among graphs belonging to the same "block" (i.e, patient), a known effect in microbiome studies. However, when applied to the brain networks, the Jaccard distance displayed an almost uniform distance between all samples and did not recover any significant clustering or grouping of patients (with a p-value associated to the analysis-of-variance test of 0.61).

**Discussion.** The Jaccard distance adjusts for graph density by including in its normalization the average sparsity of the two graphs. As such, it reflects the *amount of change with respect to the initial graph structure*. To highlight the benefits of this property, let us consider a dynamic regime in which the total number of edges stays fixed, but at each time point, each edge is replugged with probability $p$ in a previously vacant connection: the overall number of edges remains identical, but each flipped edge induces an increase in the Hamming distance of size $\frac{4}{N(N-1)}$. Hence, the average Hamming distance between $G_t$ and $G_{t+1}$ admits a closed-form expression of the type:

$$d_{\text{Hamming}}(G_t, G_{t+1}) = \frac{4p|\mathcal{E}|}{N(N-1)} = 4ps$$

where $s = |\mathcal{E}|/(N(N-1))$ is the sparsity of the original graph. By Eq. 2.3, the Jaccard distance can be written as: $d_{\text{Jaccard}}(G_t, G_{t+1}) = \frac{4p}{1+2p} = 2[1 - \frac{1}{1+2p}]$, where the later expression is a strictly increasing function of $p$. The Jaccard distance is thus independent of the sparsity and defines a one-to-one mapping between the rate of change $p$ and the observed distance. In contrast, the effect of $p$ is confounded in the Hamming distance by the influence of the sparsity.

This simple example shows that the Jaccard distance is better suited to comparing different dynamics, where the rate of edge rewiring is the main quantity of interest. Another of its advantages with respect to Hamming is that it provides a more interpretable notion of graph distances. Indeed, the Jaccard distance can be understood as the proportion of edges that have been deleted or added with respect to the total number of edges appearing in either network: a Jaccard distance close to 1 indicates an entire remodeling of the graph structure between time $t$ and $t+1$.In the microbiome study at hand, the Jaccard distance reveals more within-subject variability than Hamming distance, where the blue and red blocks in Figure 1(A) highlighted contrasted dissimilarities between graphs: here, while there exists a strong subject effect, on the whole, the almost-uniform clustermap in Figure 2(A) shows that samples within subject are still highly variable.

2.3. *Shortcomings of local approaches.*

While the Hamming and Jaccard distances provide straightforward ways of analyzing a graph's dynamics or evolution over time, such measures appear too short-sighted. Indeed, these metrics focus on the direct neighborhood of each node, and fail to capture the "bigger picture" and information on the evolution of the graph as a whole. Figure 3 shows an example where a network $G_0$ undergoes two different dynamic processes, yielding distinct graph structures with similar Hamming distances to the original. In this setup, it is possible to argue that $G_1$ and $G_2$ are more similar to each other, since the maximal path length between any two nodes is 2, whereas information percolates less rapidly across the network in the third. Conversely, from another perspective, we could also argue that we should have $d(G_1, G_3) \leq d(G_1, G_2)$, since the two first share a higher number of nodes with identical degree or since they

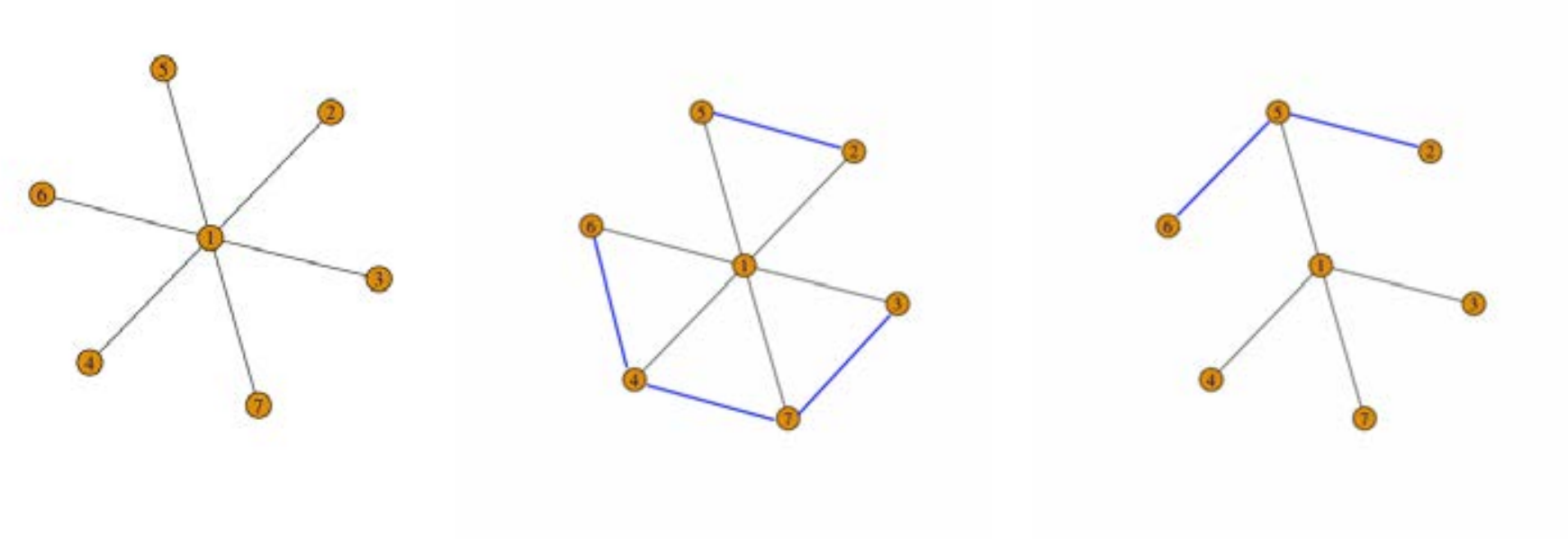

(a) Initial Graph $G_1$ (b) Perturbation of $G_1$: four edge insertions. (c) Perturbation of $G_1$: 2 edge insertions, 2 deletions.

Fig 3: Two modifications of the same initial graphs (displayed in figure 3a, such that the Hamming distance with the original is $d_H(G_1, G_2) = d_H(G_1, G_3) = \frac{4}{21}$, and the Jaccard distances are $d_J(G_1, G_2) = \frac{2}{5}$ and $d_J(G_1, G_3) = \frac{1}{4}$. The average shortest path length are 1.71 for the initial graph $G_1$, 1.51 for $G_2$ and 2 for $G_3$

have the same number of spanning trees. This example is meant to show that distances can be adapted to capture specific aspects of a network's properties. In fact, in [37], Koutra and co-authors propose to define a “good" similarity score between graphs as a score satisfying the following set of four characteristics:

1. **Edge-Importance:** modifications of the graph structure yielding disconnected components should be penalized more.
2. **Edge-Submodularity:** a specific change is more important in a graph with a few edges than in a denser graph on the same nodes.
3. **Weight Awareness:** the impact on the similarity measure increases with the weight of the modified edge.
4. **Focus awareness:** random changes in graphs are less important than targeted changes of the same extent.

These serve as guidelines and can be modified and enriched by the data analyst depending on the application at hand. The Jaccard and Hamming distance treat all edges uniformly, irrespective of their status (thus violating criterion 2 for instance).

## 3. Comparing graph structures: a spectral approach.

In order to address some of the shortcomings of the previous distances, we now turn to the class of spectral distances. Spectral distances are global measures defined using the eigenvalues of either the adjacency matrix $A$ or of some version of the Laplacian $L$. Consistent with the notation introduced in section 1, the (combinatorial) Laplacian of the graph is defined as: $L = D - A$, where $D$ is the diagonal matrix such that $D_{ii}$ is the degree of node $i$. Another popular choice consists in using the normalized Laplacian, defined as $\tilde{L} = I - D^{-1/2}AD^{-1/2}$ .

Why should eigenvalues characterize the state of a graph better? Let us first provide some intuition for this spectral approach. The eigenvalues of a graph characterize its topological structure, and in particular the way that energy or information localized at a particular node can be propagated over the graph. As such, they are related to the stability of the complex system that the graph represents. In quantum chemistry for instance, hydrocarbons are typically represented by graphs, whose adjacency matrix' eigenvalues correspond to energy levels of its electrons. In physics, the eigenvalues of the Laplacian represent the vibrational frequencies of the heat equation. The analysis of the spectral properties of a graph thus provide considerable insight into the dynamics of the system as a whole.

In this section, we brush an overview of various spectral distances. Such distances have been well studied and developed in the literature [32, 28, 3]. In [31] for instance, the authors provide an interesting review of several of such spectral distances. This deviate slightly from our original setup: spectral distances are permutation invariant and do not take into account the fact that nodes have been endowed with a particular identity. In fact, such distances can be used to compare any set of graphs, provided that they all share the same number of nodes. Spectral distances are unable to distinguish between isospectral graphs and are in fact pseudo-distances rather than actual distances. However, as the probability of having distinct graphs with identical eigenspectra quickly dwindles as the number of nodes increases, spectral distances are also viable candidates for studying the dynamics of a given complex system through time. While most current algorithms can compute eigenvalue decompositions in $O(N^3)$ steps, some computational tricks bring down this cost to $O(N^2)$ [51]– thus making this spectral approach an appealing, computationally tractable alternative for defining graph similarities [2, 22].

### 3.1. *$\ell_p$ distances on the eigenvalues.*

**Definition.** We begin by introducing a general class of versatile spectral distances. A first natural candidate for comparing two graphs based on their eigenvalue decomposition is to choose a representation of the graph (typically its adjacency matrix, combinatorial or normalized Laplacian,etc.) and to simply consider the $\ell_p$ distance between functions of their eigenspectra.

For any (almost everywhere) differentiable function of the graph's eigenvalues $\lambda_0 \leq \lambda_1 \leq \cdots \leq \lambda_{N-1}$, we can write:

$$d(G, \tilde{G})^p = \sum_{i=0}^{N-1} |f(\lambda_i + \epsilon_i) - f(\lambda_i)|^p \approx \sum_{i=0}^{N-1} |f'(\lambda_i)|^p |\epsilon_i|^p \tag{3.1}$$

An important step thus consists in picking an adequate representation of the graph: how to decide between using the graph's adjacency matrix, its Laplacian or normalized Laplacian? These representations and the relationship between their eigenvalues and properties of the graph (average degree, Cheeger constant, etc.) have been well investigated in the literature [12], yet no consensus as to which representation yields more accurate results for comparing graphs has been established. To try and resolve this issue, we suggest the following guidelines:

- *leveraging the representation's physical interpretation.* As underlined in the introductory paragraph of this section, both the eigenvalues of the Laplacian and those of the adjacency matrix can be related to physical properties of a system and can thus be considered as characteristics of its states. Whenever such a physical interpretation exists, a good choice thus lies in the selection of the corresponding representation.
- *opting for the most robust alternative.* The adjacency matrix does not down-weight any changes and treats all nodes equivalently. On the other hand, the eigenspectrum of the Laplacian accounts for the degree of the nodes and is known to be robust to most perturbations: a "small" perturbation of the graph –that is, a perturbation that has very little impact on the graph's overall connectivity– will only induce a small change in the eigenvalues [49], thus making them a more attractive alternative for comparing graph structures.
- *choosing a stable representation.* The literature remains divided on which version of the Laplacian to pick. However, the eigenvalues of the normalized Laplacian are bounded between 0 and 2, making it a somehow more stable and preferable representation.

**Application.** Let's look at the spectral distances on our microbiome and brain data. Figure 4 shows the results for the $\ell_2$ distance using two different functions of the combinatorial Laplacian eigenspectrum in Eq. 3.1 on our datasets: the low-pass filters with randomly chosen parameters $f(\lambda) = e^{-0.1\lambda}$ (Microbiome, top row of Figure 4) and $f(\lambda) = e^{-1.2\lambda}$ (rs-fMRI, bottom row of Figure 4). The first interesting observation that we make is that these distances produce different results than the previous distances. In particular, we note that for $f(\lambda) = e^{-0.1\lambda}$, the nearest-neighbor metagraph is significantly associated with the treatment stages labels: the p-value associated to the analysis-of-variance test yields a value of 0.025, which is confirmed by the MDS projections (Figure 4 B) showing a clear grouping of the graphs per subject. This effect subsides as the scaling value increases while its association with the treatment stage becomes predominant. For $f(\lambda) = e^{-1.2\lambda}$, the analysis-of-variance test (with stages as labels) yields a p-value of 0.015. We also note that this distance is the only one which recovers some meaningful associations between the nn-metagraph of the fMRI data and the age under dependency (Fig.4D). We also note that the choice of the representation matters: the fMRI dataset analyzed through the scope of spectral distances based on the adjacency matrix failed to reveal any significant effect. Note that in the microbiome example and the synthetic experiments detailed in section 5, the choice of one representation over another was mitigated, with both the Laplacian and the adjacency matrix yielding comparable results. Moreover, as underlined above, the choice of the function itself can lead to the discovery of different effects.

To understand this phenomenon, we build upon the signal processing analogy developed in [46]. In tha paper, Shuman and co-authors show that the eigenvalues of the Laplacian can be interpreted as the analog of a signal's frequencies in the temporal domain. Low eigenvalues and their corresponding eigenvectors are analogous to slowly-varying low-frequency signals over the graph: if two vertices are connected by an edge with a large weight, the values of the eigenvector at those locations are likely to be similar. By contrast, the eigenvectors associated to high eigenvalues vary more rapidly across edges [47, 52]. Hence, "low" eigenvectors encapsulate local information about the structure of the graph (yielding results akin to the Jaccard distance in the microbiome example) while higher values of $\alpha$s cover a larger portion of the spectrum and allow the incorporation of more global information.

**Discussion.** We continue upon the signal processing analogy to find an appropriate choice of the function $f$:

- If the goal of the analysis is to capture *the importance of the changes in the connectivity of the overall graph structure*, the distance should put more emphasis on the first eigenvectors. An adequate choice for $f$ would be thus to select $f$ to act as a low-pass filter: putting more weight on changes occurring in small eigenvalues, and discounting the effect of changes at the higher end of the spectrum. The strength of

(A)

(B)

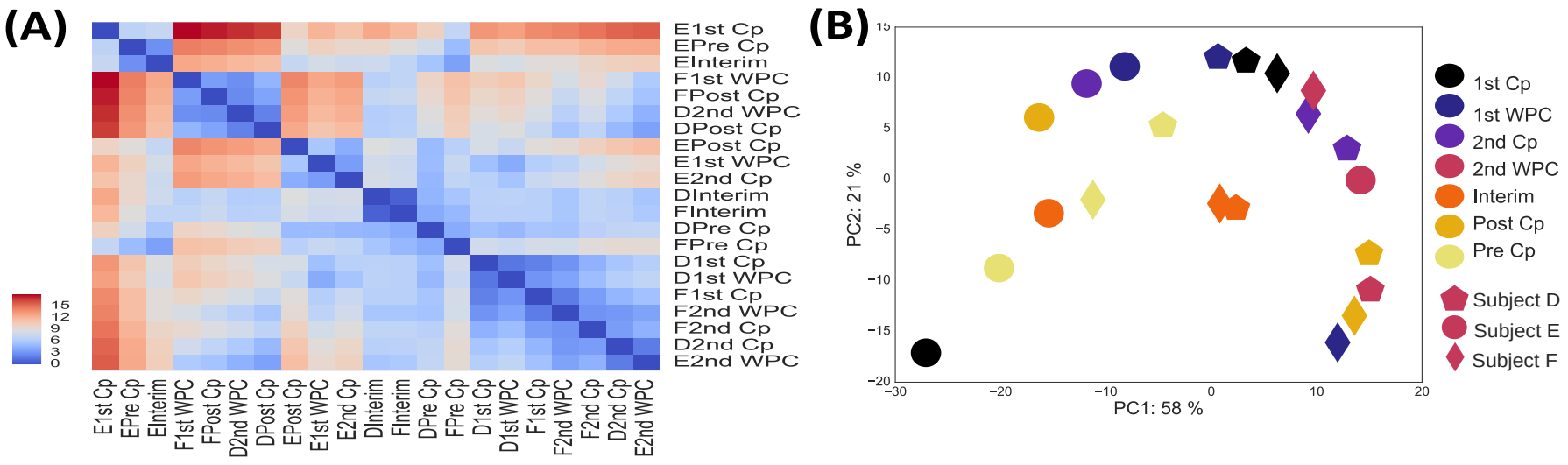

(C)

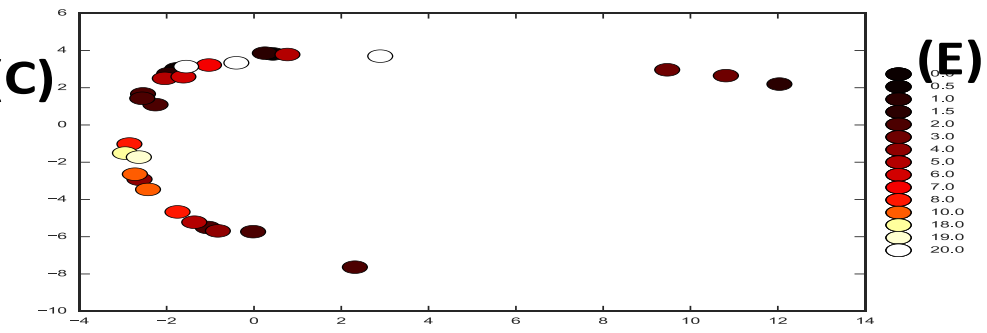

(E)

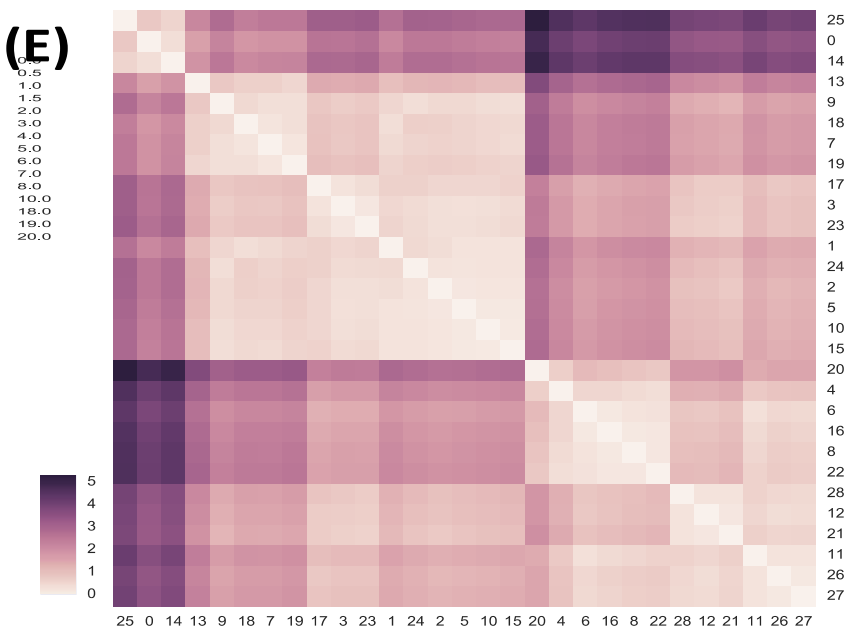

(D)

| Dataset and Labels | FR p-values (nn-metagraph) | Analysis-of-variance Test |
|---|---|---|
| **Relman- Microbiome** (3 classes, one per subject) | 0.640 | 0.306 |
| **Relman- Microbiome** (7 classes, one per treatment stage) | 0.047 | 0.0148 |
| **Kelly- fMRI dataset** (continuous labels, years of dependency) | 0.077 | 0.296 |

Fig 4: Application of $\ell_2$ spectral distances using two functions of the Laplacian eigenspectra in Eq. 3.1: low-pass filters $f(\lambda) = \epsilon^{-0.1\lambda}$ on the Microbiome (**top row**) and $f(\lambda) = \epsilon^{-1.2\lambda}$ on the fMRI dataset( **bottom row**). Clustermap of the corresponding distances between bacterial graphs **(A)**/ brain connectomes **(E)** . MDS projection of the bacterial **(B)**/ fMRI **(C)** graphs on the first two principal axes. Colors denote treatment phases/ years of dependency. Pvalue of the FR-test for the 1-nn metagraph across the different datasets, for the low-pass filter $f(\lambda) = \epsilon^{-1.2\lambda}$ **(D)**.

the modulation of the eigenvalues by the filter depends on the analysis. For instance, taking $f : x \to e^{-\alpha x/\lambda_3}$ ensures associating a weight of at most $\frac{\alpha}{\lambda_3}\epsilon^{-\alpha}$ in Eq. 3.1 on changes in eigenvalues greater or equal to $\lambda_3$. In the case where $\lambda_2 << \lambda_3$, this gives more importance to changes occurring in $\lambda_2$. In our microbiome study, as previously highlighted, we recover more structure in the dataset by focusing on the lower part of the spectrum (Figure 4) and discarding the higher frequency, "noisier" eigenvalues.

- Supposing that one is interested in the *overall change in the "graph's frequencies" at every level* induced by the perturbation, one might actually prefer to take a function that would not discriminate against any value of the eigenfunction, but simply look at the amplitude of the change in eigenvalue. In that case, $f$ could simply be taken to be the identity.

This section has shown the possibility of crafting a distance based on the Laplacian or the adjacency matrix eigenspectrum, tailored to the requirements and objectives of the analysis. However, choosing "an optimal" kernel function for the problem at hand requires domain knowledge or additional insight into the problem – thus requiring more thought than the straightforward Hamming distance.

### 3.2. *Spanning tree similarities.*

**Definition.** Inspired by A. Kelmans [34, 35], who characterized transformations by their "ability to destroy", we now introduce a similarity which reflects the number of spanning trees that are destroyed or created by the transformation of one graph to another.

The Matrix-Tree theorem provides us with a convenient way of computing the number of spanning trees for a connected graph: denoting $0 = \lambda_0 < \lambda_1 \leq \cdots \lambda_{N-1}$ the eigenvalues of the Graph Laplacian $L = D - A$, we have:

$$\mathcal{T}_G = N_{\text{Spanning tree of } G} = \frac{1}{N} \prod_{i=1}^{N-1} \lambda_i$$

A dissimilarity between two graphs $G$ and $\tilde{G}$ can be defined by comparing the quantities:

$$d_{ST}(G, \tilde{G}) = \left| \log(\mathcal{T}_G) - \log(\mathcal{T}_{\tilde{G}}) \right| \tag{3.2}$$

On an intuitive level, spanning trees are a reflection of the graph's interconnectedness and robustness to change: to draw an analogy with electric current, this amounts to quantifying the effect of one edge deletion on the impedance of the system: how easily does the current still manage to flow?

**Application.** In this case, the results obtained using the spanning tree dissimilarity (denoted as ST dissimilarity in the rest of the text) are comparable to the results provided by the low-pass filter spectral distance described in the previous subsection. We also observe an interesting phenomenon: the nearest-neighbor metagraph induced by the ST dissimilarity on the microbiome data is fairly consistent with the treatment stages (with an associated FR-pvalue of 0.158), but, as we increase the number of neighbors, this effect becomes rapidly insignificant. However, these later $k$-nearest neighbor graphs are significantly associated to the subject labels (below the 5% threshold). This effect is confirmed by the analysis-of-variance test described in section 2.2, which yields a significant p-value of 0.035.This indicates that the ST dissimilarity does capture both similarities between treatment phases as well as across subjects. This effect can be visualized in Figure 5B: the MDS projections of the microbiome graphs along the first 2 principal components follow a curve (which is indicative of a gradient in higher dimensions), along which points belonging to the same subject seem relatively close. Similarly as before, the Spanning tree distance recovers some structure in the fMRI datasets (lighter blocks along the diagonal in Figure 5D), although there is no evidence that these clusters are associated with the time under dependency.
**Discussion.** Suppose that graph $G$ undergoes a "”small” perturbation, yielding a new graph $\tilde{G} = \mathcal{T}(G)$. We know that the eigenvalues of $\tilde{G}$ can be written as a perturbed version of the eigenvalues of $G$, that is:

$$\forall i, \quad \tilde{\lambda}_i = \lambda_i + \epsilon_i$$

Hence, we can write:
(3.3)

$$\tilde{\mathcal{T}}_G = N_{\text{Spanning tree of } \tilde{G}} = \frac{1}{N}\prod_{i=1}^{N-1}\tilde{\lambda}_i = \mathcal{T}_G \times [1 + \sum_{i=1}^{N-1}\frac{\epsilon_i}{\lambda_i} + \sum_{i,j=1}^{N-1}\frac{\epsilon_i\epsilon_j}{\lambda_i\lambda_j} + \cdots]$$

Combining 3.3 and 3.2 yields:

$$d_{ST}(G, \tilde{G}) = |\log(1 + \sum_{i=1}^{N-1}\frac{\epsilon_i}{\lambda_i} + \sum_{i,j=1}^{N-1}\frac{\epsilon_i\epsilon_j}{\lambda_i\lambda_j} + \cdots)| \tag{3.4}$$

The impact of the change is thus inversely proportional to the value of the eigenvalues. This is an attractive property for weakly connected graphs (i.e, that have small $\lambda_1$), where the addition or deletion of a critical edge can have a huge impact on the graph's overall connectivity. Conversely, changes in larger eigenvalues have less impact: to continue with the temporal frequency analogy drawn in the previous section, this similarity automatically discounts changes that are related to noise, and accentuates the impact of changes on low eigenvalues which are considered to be more reflective of the graph's structure. We emphasize that, once again, this defines a pseudo-distance between eigenspectra (or dissimilarity score between graphs), rather than a distance on the graphs themselves. This effect is shown in Figure 5, which exhibits results close to the low-pass filter approach developed in the previous subsection. The advantage of the ST dissimilarity is that it does not require the specification of a particular ad-hoc low-pass kernel on the eigenspectrum. However, this does come at an increased price in terms of the variability of the results: because the effect of perturbations is measured with respect to the inverse of the eigenvalues (Eq. 3.4), this distance is less stable than the low-pass filter spectral distance. We will study this in more depth in our synthetic experiments in section 5.

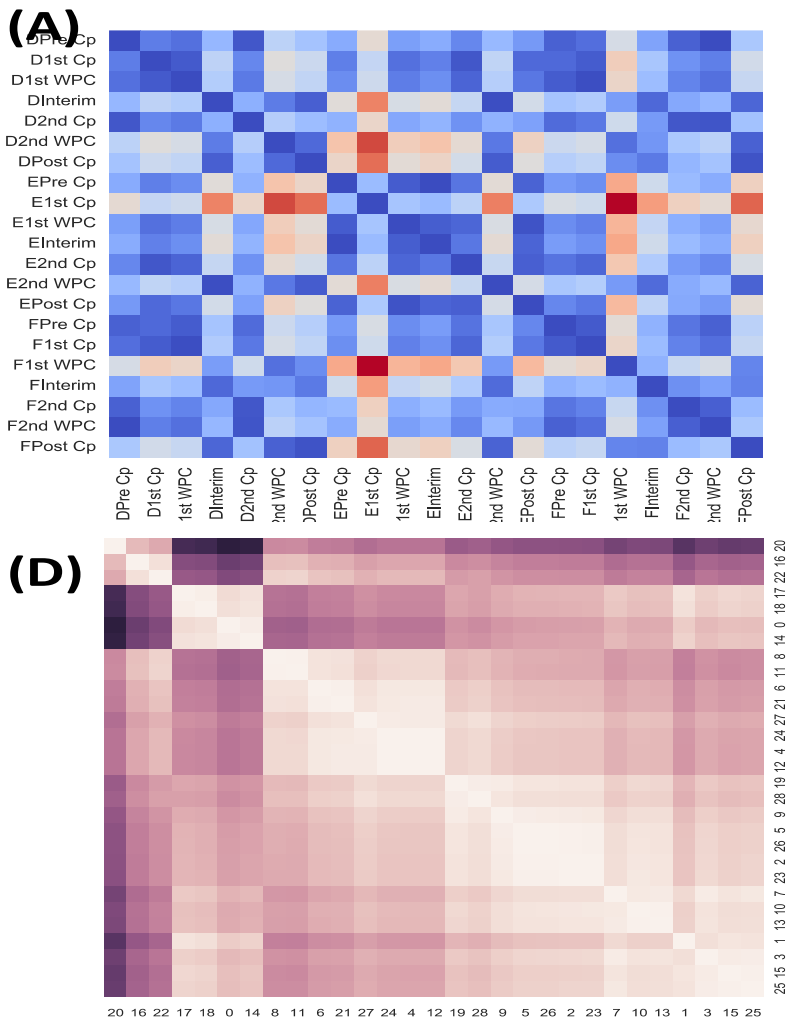

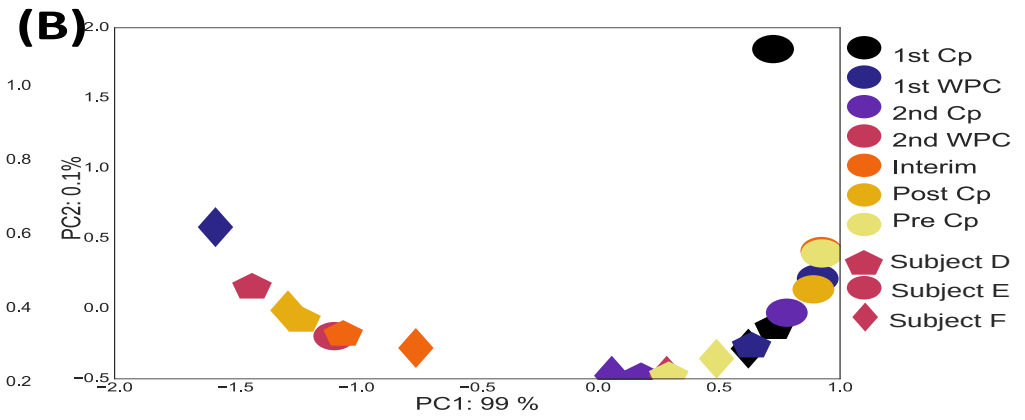

(C)

| Dataset and Labels | FR p-values (3nn-metagraph) | Analysis of Variance Test |
|---|---|---|
| **Relman- Microbiome** (3 classes, one per subject) | 0.041 | 0.037 |
| **Relman- Microbiome** (7 classes, one per treatment stage) | 0.316 | 0.349 |
| **Kelly- fMRI dataset** (continuous labels, years of dependency) | 0.865 | 0.225 |

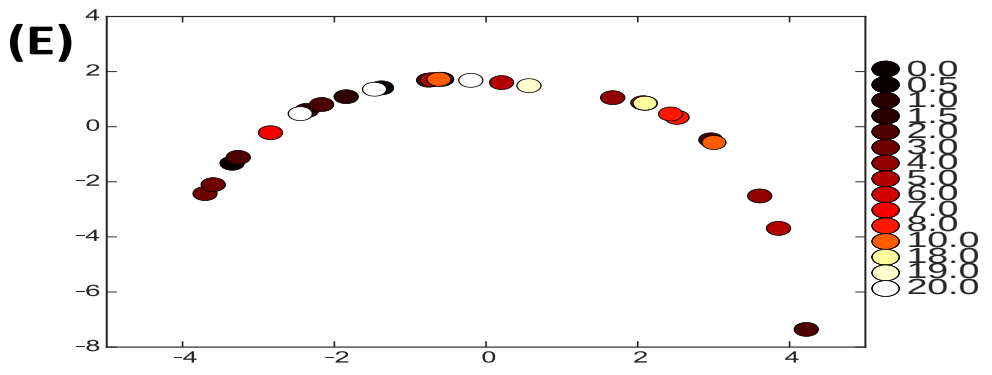

Fig 5: Application of the Spanning tree dissimilarity. **(Top) Microbiome Data/(Bottom) fMRI Data.** Heatmap of the corresponding dissimilarity between Kendall-correlation-based bacterial graphs **(A)**. MDS projection of the bacterial graphs on the first two principal axes **(B)**. Colors denote treatment phases, and shapes represent different subjects. p-values associated to Friedman-Rafsky test of the consistency of the 3-nn metagraph with the labeling of the nodes and analysis-of-variance test **(C)** Clustermap for the fMRI data **(D)**. MDS projections of the brain connectomes on the first two principal components **(E)**.

3.3. *Distances based on the eigenspectrum distributions.*

3.3.1. *General framework.* Rather than focusing on the graph's eigenspectra, another alternative proposed by [31, 22] considers *continuous spectral distributions*. The continuous spectral distribution is obtained from each graph by computing the graph's eigenvalues and considering a kernelized version of its eigenvalue distribution. For a Gaussian kernel, the spectral distribution is defined as:

$$\rho_G(x) = \frac{1}{n}\sum_{i=0}^{n-1} \frac{1}{\sqrt{2\pi\sigma^2}} e^{-\frac{(x-\lambda_i)^2}{2\sigma^2}}$$

A pseudo-distance between graphs is based on the distance between spectrum distributions, which, in the case of the $\ell_1$-distance used in [22], yields the following expression:

$$d(G, \tilde{G}) = \int \big|\rho_G(x) - \rho_{G'}(x)\big| dx.$$

In their 2016 article [22], Gu and co-authors show that, in the limit of an infinite number of nodes, these distances have the added benefit of distinguishing between different types of graphs (Erdös-Rényi vs Preferential Attachment, etc). As such, these distances are able to recognize important geometrical information in the overall graph structure. We now investigate a variant of such a class of distances. Proposed in [26], it has been shown to exhibit interesting properties [30]: the Ipsen-Mikhailov distance.

3.3.2. *Definition of the IM distance.*

**Definition.** First introduced by Ipsen[26] for graph reconstruction and later extended to the broader "graph-comparison" problem by Jurman and co-authors [29, 30, 31], the Ipsen-Mikhailov distance is a spectral measure which relates a network on $N$ nodes to a system with $N$ molecules connected by elastic strings. The connections are dictated by the graph's adjacency matrix $A$ and the system can thus be described by a set of $N$ equilibrium equations:

$$\frac{\partial^2 x_i}{\partial t^2} + \sum_{j \neq i} A_{ij}(x_i - x_j) = 0$$

In this setting, the eigenvalues of the Laplacian matrix of the network are interpreted as the squares of the vibrational frequencies $\omega_i$ of the system: $\lambda_i = \omega_i^2$ with $\lambda_0 = \omega_0 = 0$.

The Ipsen-Mikhailov distance characterizes the difference between two graphs by comparing *their spectral densities*, rather than the raw eigenvalues themselves. The spectral density of a graph is defined as the sum of Lorenz distributions:

$$\rho(\omega, \gamma) = K \sum_{i=1}^{N-1} \frac{\gamma}{\gamma^2 + (\omega - \omega_i)^2}$$

where $\gamma$ is a parameter common to all vibrational frequencies that we will have to determine, and $K$ is the normalization constant defined such that: $\int_0^\infty \rho(\omega, \gamma) d\omega = 1$. This spectral distance between two graphs $A$ and $B$ is defined as:

$$\epsilon_\gamma(A, B) = \sqrt{\int_0^\infty [\rho_A(\omega, \gamma) - \rho_B(\omega, \gamma)]^2 d\omega} \tag{3.5}$$

The latter expression depends on the choice of the scale parameter $\gamma$. Jurman and co-authors [31] set $\gamma = \bar{\gamma}$ as the unique solution of:

$$\epsilon_{\bar{\gamma}}(\mathcal{E}_N, \mathcal{F}_N) = 1$$

So the IM distance is bounded between 0 and 1 and its upper bound is attained only for $\{A, B\} = \{\mathcal{E}_N, \mathcal{F}_N\}$ where $\mathcal{E}_N$ denotes the empty graph and $\mathcal{F}_N$ the complete graph on $N$ nodes. In Appendix E, we investigate a closed form formula for these parameters.

3.3.3. *The Hamming-Ipsen-Mikhailov distance .*

**Definition.** So far, none of these spectral distances have used the fact that particular nodes can be matched. There is no way of discriminating changes (that is, emphasizing changes in areas of the graph deemed important to the analyst), or of accounting for rare - but existing- isospectral graphs.
To bridge the two approaches, Jurman and al [30] propose a distance that is a weighted linear combination of the Ipsen-Mikhailov and the normalized Hamming.

$$d^{\xi}_{HIM} = \frac{1}{\sqrt{1+\xi}}\sqrt{IM^2 + \xi H^2}$$

**Application.** The results of the microbiome analysis carried out with this distance are displayed in Figure 8. We note that the improvement with respect to the Ipsen-Mikahilov distance is only marginal.
**Discussion.** This distance benefits from the advantages of both the Hamming and the Ipsen-Mikhailov distances by combining local and global information. Note that, since it is a linear combination of a distance with a non-negative quantity, this defines a proper distance between graphs. The parameter $\xi$ provides additional flexibility to the metric by allowing to favor one type of information over another. However, empirically, we have observed this distance to be computationally expensive, and thus difficult to apply to the study of large graphs and/or large datasets.
**Application.** Figure 6 shows the results of the analysis using the HIM distance on our microbiome study. The MDS projection (Figure 6A) seem to highlight a similarity between graphs corresponding to the same treatment. The Friedman-Rafsky test on the minimum spanning tree with the treatment phases as labels is significant, with a p-value of 0.00048. This is further confirmed by the analysis of variance test described in section 2.2 with the stages as labels, yielding a pvalue of $p = 10^{-5}$. As also shown by Figure 6C, the HIM is able to make the best of both the Hamming and spectral distances, and is thus able to spot more structure in the datasets. Overall, because these spectral distances are "unlocalized" and make no use of the nodes' identities, they are suited to the comparison of graphs' overall structure without any prior on where “critical” changes occur in the spectrum. On an aside note, both the IM and HIM distances were the lengthiest to compute – perhaps restricting their scope of use to the comparison of small sets of reasonably-sized graphs.

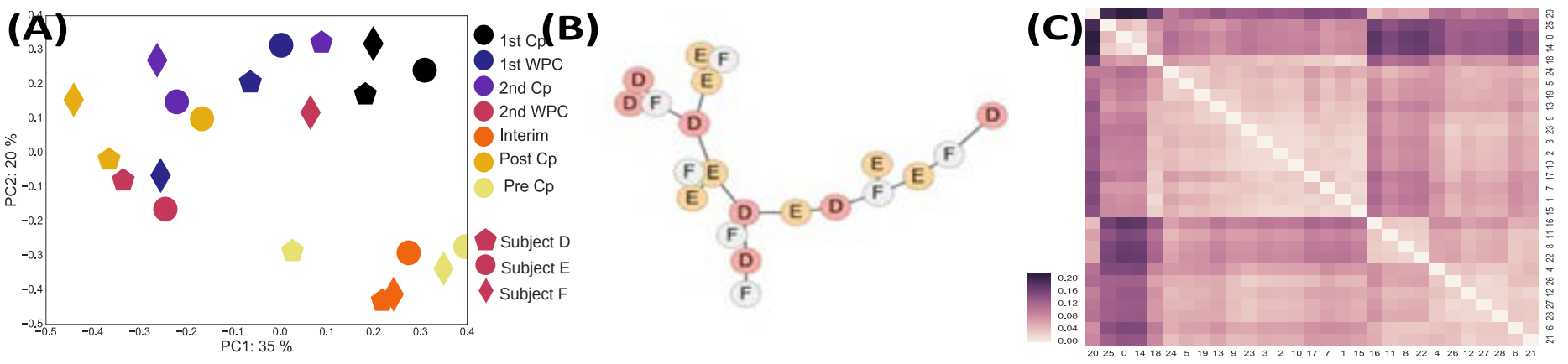

Fig 6: Application of the Hamming-Ipsen-Mikhailov distance. MDS projection of the bacterial graphs on the first two principal axes **(A)** . Colors denote treatment phases, and shapes represent different subjects. Minimum Spanning Tree induced on the bacterial graphs by the HIM distance **(B)**. Application of the HIM distance to the fMRI data set: clustermap of the different distances between connectomes **(C)**.

### 3.4. *The Polynomial Approach.*

**Definition and motivation.** The previous spectral distances all shared a common problem: they require an explicit computation of the graph's eigenvalues – which, computational tricks aside, still generally has complexity $O(N^3)$ – and is sensitive to global properties of the graph (as captured by the eigenvalues). Structural distances (Hamming and Jaccard) however, were too short-sighted and concentrated on changes in each node's direct neighborhood. Another interesting type of distances would thus be at an intermediate scale, and compare changes in local neighborhoods. For instance, changes in sparse regions of the graphs might be more informative than perturbations in very dense ones. Following Koutra and co-authors's [37] proposed guidelines for distance selection, a "good" similarity score should be able to capture such nuances and attribute more weight to changes in areas of the graphs deemed more critical by the data analyst.

In this new setup, a possible solution is to work directly with the powers of the graphs' adjacency matrix $A^k$. Indeed, the powers of the adjacency matrix relate directly to a graph's local topology through the coefficients $A^k_{ij}$, which corresponds to the number of paths (possibly with cycles) that start at $i$ and arrive at $j$ in $k$ hops. Hence, by design, these powers are inherently local. The coefficients $A^k_{ij}$ can be thought of as a characterization of the connectivity between two nodes with respect to the $k$-hop neighborhoods: nodes $i$ and $j$ at distance greater than $k$ hops have connectivity index $A^k_{ij} = 0$, whereas nodes within each other's $k$ -hop neighborhood will typically have high connectivity index $A^k_{ij}$ if the neighborhood is dense, and lower $A^k_{ij}$ if the region is sparse. As such, the powers of the adjacency matrix seem to offer an attractive starting point to quantify changes on the mesoscale.

Typically, for each neighborhood (centered around a node $a$), perturbations should be assigned weights that are monotonically decreasing functions of the distance: a perturbation has higher impact in the local neighborhood if it is closer to the center than the periphery. In this spirit, denoting as $A = Q\Lambda_A Q^T$ the eigenvalue decomposition of the adjacency matrix $A$ of a given graph, a proposed similarity score is defined with a polynomial $P(x) = x + \frac{1}{(N-1)^\alpha}x^2 + \cdots + \frac{1}{(N-1)^{\alpha(K-1)}}x^K$ of the adjacency:

$$P(A) = QWQ^T$$

where $W = \Lambda_A + \frac{1}{(N-1)^\alpha}\Lambda_A^2 + \cdots + \frac{1}{(N-1)^{\alpha(K-1)}}\Lambda_A^K$.

The distance between two graphs $G_1$ and $G_2$ can simply be computed by comparing the polynomials of their associated adjacency matrices $A_1$ and $A_2$:

$$d_{\text{pol1}}(G_1, G_2) = \frac{1}{N^2}||P(A_1) - P(A_2)||_{2,2} \tag{3.6}$$

In a way, this distance is a straightforward extension of the Hamming distance to the mesoscale: rather than looking at perturbations at the atomic level – counting the number of removed and inserted edges without assessing the effect of the perturbation on the overall structure, this polynomial distance compares neighborhoods of larger sizes and thus attempt to capture the effect of perturbation at an intermediate scale. The weighting factor $\alpha$ is a way of discounting "peripheral" changes in neighborhoods of larger sizes with respect to neighborhoods of smaller size. We note that Eq. 3.6 is just a proposed class of polynomial distances, but this set of distance can be more broadly customized to a specific problem at hand, including domain knowledge to choose the size of the neighborhood, etc.

**Application.** Figure 7 shows the application of the polynomial distance to the microbiome data. Similar to the Hamming distance, the polynomial distance does detect significant similar dynamics across subjects (closely matching curves in Figures 7C). However, in this case, the polynomial approach seems a weak compromise between structural and spectral distances, and does not benefit from any of their advantages: the polynomial distance is neither significantly associated to states or subjects (as per the associated Friedman-Rafsky and analysis-of-variance type tests).

**Comparison of polynomial, spectral and structural distances.** The main advantage of the polynomial distances over the Hamming and Jaccard distances is that the former takes into account the properties of each node

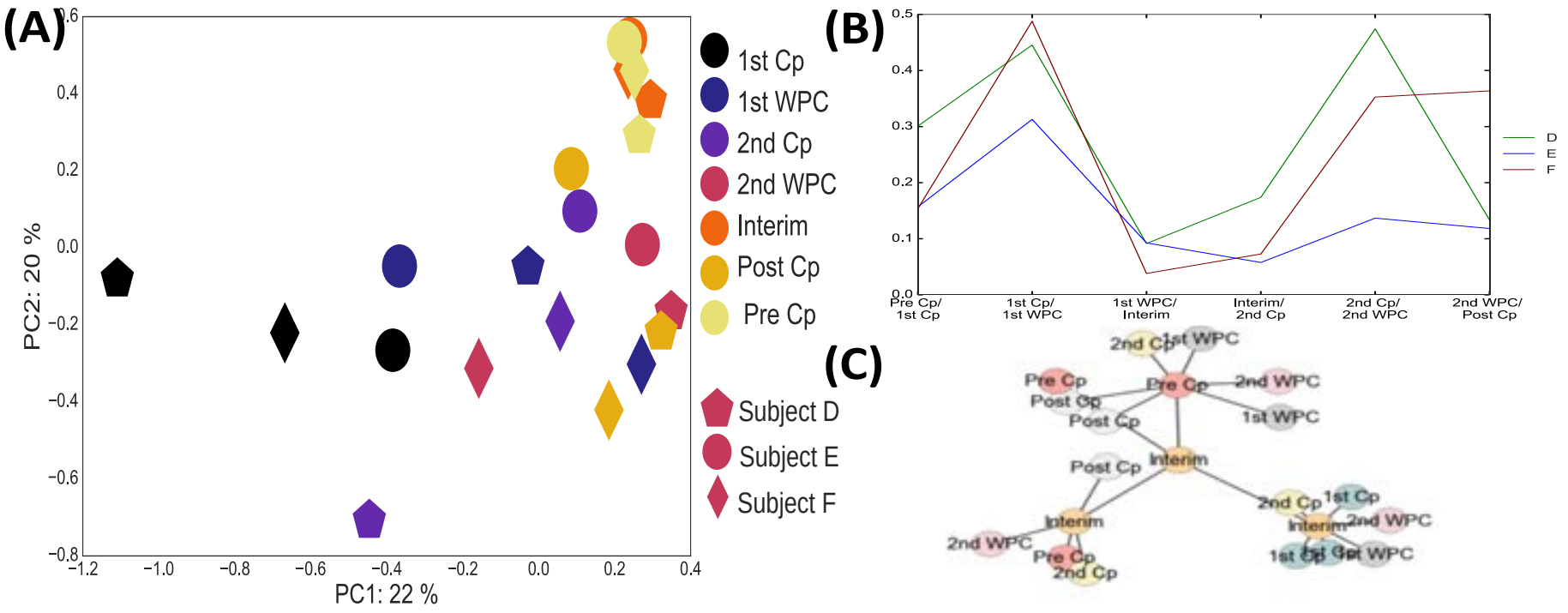

Fig 7: Application of the polynomial dissimilarity to the microbiome bacterial graphs, for $K = 3, \alpha = 0.9$. Heatmap of the corresponding dissimilarity **(A)** . MDS projection of the bacterial graphs **(B)** on the first two principal axes. Colors denote treatment phases, and shapes represent different subjects. Plots of the consecutive distances between bacterial graphs **(C)**.

– past each node's immediate neighborhood. Indeed, it is possible to show that the effect of a perturbation (that is, the addition of one edge) on the graph can be directly related to properties of the graph at a higher order than simply the one-hop neighborhood. By construction, polynomials of order $k$ reflect the effect of the perturbation on $k$-hop neighborhoods. They can be expressed in terms of polynomials of the degree of the nodes of the added edge, as well as the size of the intersection of their neighborhoods (up to size $k$). Hence, while its form (powers of the Laplacian) make it an intrinsically local distance, the polynomial distance is a first step towards bridging purely structural and spectral distances, by extending the Hamming distance to neighborhoods of greater depth. However, this might come at an increased price: the real-application studies have shown this distance to be blurred by both the perturbation of the organizational structure of the microbiome from one phase to the next and the variability of the bacteria across subjects.

## 4. Quantifying change at the mesoscale.

Most of the distances described in the previous sections have failed our objective in some way or other: structural distances have proven to be too "local" and agnostic to perturbations' effects on a given complex system's organization as a whole. Spectral distances are too global and fail to use the information captured in the nodes' identities. Polynomial distances – which quantify changes with respect to the $k$-hop neighborhoods – on the other hand have shown promising properties in both real and synthetic experiments: in the case where nodes' identities hold some insightful information, this extension of standard structural metrics seem to have brought a solution, trading off between the locality of the changes and their impact on the organization of the system as a whole. This indicates that considerable insight can be gained by comparing graphs at this intermediate "neighborhood" scale. This approach thus calls for the need for characterizing topological properties of these neighborhoods. In this section, we investigate graph comparison through a "glocal" lense (borrowing an expression from [30]), extending the class of mesoscale polynomial distances introduced in section 3.4 by suggesting two alternative characterizations of neighborhoods' topological properties.

### 4.1. *Quantifying interactions: connectivity-based distances.*

We begin with a simple intuitive distance based on some measure of nodes' pairwise interactions. Indeed, as previously underlined, we want a distance that: (a) preserves information about each node's identity and (b) incorporates information characterizing nodes by their relationship to the whole graphs, rather than uniquely with respect to their direct neighbors. A general framework is to consider the set of graph dissimilarities defined as:

$$d_{\text{centrality}}(G_t, G_{t+1}) = \Big( \sum_{i=1}^{n} \sum_{j=1}^{n} (s_{ij}^{(t+1)} - s_{ij}^{(t)})^p \Big)^{1/p} \tag{4.1}$$

where $s_{ij}^t$ is some measure of the interaction or affinity between nodes $i$ and $j$ in graph $G_t$. This dissimilarity metric thus quantifies how much the different interactions have changed from one graph to the other. This approach satisfies our constraints: it is both local and respects nodes' identities while accounting for the whole graph structure by summing over all pairwise "interaction" scores.

In the simplest, most intuitive case, we can simplify this expression by using centrality measures. Indeed, centrality measures (betweenness, harmonic, etc.) can typically be used to characterize them as either belonging to part of the core or the periphery of the graph, and thus encode global topological information on the status of node within the graph. These metrics are thus natural candidates to characterize 'mesoscopic' changes. More formally, in this setting, denoting $c_i^{(t)}$ as the betweenness-centrality of each node $i$ in the graph at time $t$, one defines a distance between two graphs $G_t$ and $G_{t+1}$ as:

$$d_{\text{centrality}}(G_t, G_{t+1}) = \sqrt{\sum_{i=1}^{n} (c_i^{(t+1)} - c_i^{(t)})^2} \tag{4.2}$$

One of the positive aspects of this metric is that centrality measures are "integrated" quantities; measuring the number of paths that typically pass through a given node. As such, similarly to eigenvalues, these metrics are more robust to small perturbations in the graph structure than the Hamming distance. Moreover, each "drift" measure in Eq. 4.2 is interpretable: a change in centrality can be understood as a drift of the node away from (or towards) the core of the network. However, the problems associated with this approach are two-fold. First of all, one has to choose a single "good" centrality measure (harmonic, betweenness, etc.), which may require domain-knowledge, since this captures a specific aspect of the network's evolution. Moreover, the computation of betweenness centrality on unweighted graphs typically requires algorithms with complexity $O(|\mathcal{V}||\mathcal{E}|)$. This approach is thus unfortunately, like the IM distance, difficult to extend to larger graphs.

In order to make this approach more tractable, recent work has proposed using approximation algorithms to compute alternative interaction metrics in Eq.4.1. For instance, in [43] Papadimitriou and co-authors suggest five different scalable similarities. In [37], Koutrai and co-authors propose a low-dimensional approximation of these scores based on loopy belief propagation algorithm – yielding a method (DeltaCon), able to approximate Eq.4.1 with a computational complexity linear on the number of edges in the graphs.

4.2. *Heat spectral wavelets.*

Another alternative is to derive characterizations of each node's topological properties through a signal processing approach: the values of the nodes constitute a signal over the graph, which can be filtered by modulating the graph's spectrum. This yields a different similarity than in the eigenvalue-based setting: whereas in the previous section, eigenvalue distances simply computed a distance between the modulation of two graphs' eigenvalues, here, the eigenvalues are modulated and combined with their respective eigenvectors to yield a "filtered" representation of the graph's signal. Such an approach could follow work initiatially done by Monning and co-authors, who, in their recent 2016 paper [42], build upon the DeltaCon similarity to create a (proper) distance between graphs: they introduce the *Resistance Perturbation* index, a metric based on the eigenvalues and eigenvectors of a modified version of the graph Laplacian. In this section, we focus on a closer analogy to signal processing and use recent work in the graph signal processing literature to derive such characterizations. In this subsection we focus on an approach inspired by [15] for the purpose of structural role identification. In that paper, inspired by the emerging field of graph signal processing [46], the authors suggest using heat spectral wavelets to characterize each node's local topology for the purpose of structural role identification. To use a concrete analogy, this method operates in a way similar to sonar detection: each node probes the network by diffusing a heat wavelet, and the way that the network responds to each of these probes – that is, the different heat prints that are obtained for each node – is taken as a signature for each of the nodes' topological neighborhoods.

More formally, denoting $L = U\Lambda U^T$ as the Laplacian's eigenvalue decomposition, where $0 = \lambda_0 \leq \lambda_1 \leq \cdots \leq \lambda_{N-1}$, the heat scaling-wavelet [47] $\Psi^{(\tau)}_{\cdot,a}$ centered at node $a$ with scale $\tau$ is defined as the column vector of the matrix $Ue^{-\tau\Lambda}U^T$:

$$\Psi^{(\tau)}_{\cdot,a} = Ue^{-\tau\Lambda}U^T\delta_a \implies \forall m, \quad \Psi^{(\tau)}_{m,a} = \delta_m^T Ue^{-\tau\Lambda}U^T\delta_a = \sum_{j=0}^{N-1} e^{-s\lambda_j} U_{aj}U_{mj}$$

where $\delta_m$ is the indicator vector associated to node $m$, and $e^{-\tau\Lambda}$ is the diagonal matrix $\mathrm{Diag}(e^{-\tau\lambda_0}, e^{-\tau\lambda_1}, \cdots, e^{-\tau\lambda_{N-1}})$. In [15], the authors propose to define a structural signature for each node as the unordered set of coefficients:

$$\chi_a = \{\Psi^{(\tau)}_{m,a}\}. \tag{4.3}$$

By comparing these distributions, one captures information on the connectedness and centrality of each node within the network, thereby providing a way to encompass in a Euclidean vector all the necessary information to characterize nodes' topological status within the graph. Moreover, in order to compute these wavelets in a tractable fashion that extends to large graphs, Hammond and co-authors [24] suggest the use of Chebychev polynomial approximations.The cost of computing the wavelet transforms becomes simply $O(K|\mathcal{E}|)$–making spectral wavelets an attractive approach for characterizing structural roles. While these signatures were initially devised to detect structural similarities across a network, they can also be employed to characterize similarities across a set of aligned graphs. In this setting, network similarity between graphs $G_t$ and $G_s$ is defined by comparing each node's topological signature in $G_t$ with its counterpart in $G_s$. Here, a large dissimilarity between graphs indicates either an important "volume" of change (as in the Hamming distance) or that some nodes have undergone important topological changes. It thus captures changes at both the fine and intermediary scales.

More formally, this dissimilarity between graphs amounts to an $\ell_2$ distance between each node's structural embedding:

$$
(4.4) \quad \begin{aligned} d(G_t, G_{t+1}) &= \frac{1}{N}\sum_{a\in\mathcal{V}} ||r_a^{(t)} - r_a^{(t+1)}||_2 = \frac{1}{N}\sum_{a\in\mathcal{V}} \delta_a \Delta^T \Delta \delta_a \\ \implies d(G_t, G_{t+1}) &= \frac{1}{N}\text{Tr}[\Delta^T \Delta] \end{aligned}
$$

where $\Delta = U_t e^{-\tau\Lambda_t} U_t^T - \tilde{U_{t+1}} e^{-\tau\Lambda_{t+1}} U_{t+1}^T$.

To formalize the link with the previous subsection, we argue that these wavelet coefficients are in fact robust integrated centrality scores. Indeed heat kernels can be understood as a robust *page rank score* at each node [10]. By design, the heat kernel integrates over the neighborhoods (the size of which depends on the scale of the kernel) and is thus less sensitive to small perturbations. Hence, these wavelet coefficients provide a tractable alternative to the centrality measures proposed in section 4.1. As such, they benefit from these measures' interpretability, while being generalizable to larger networks.

**Discussion.** We summarize the advantages of this method as follows.

- **tractability**: as already noted, the cost of computing the wavelets via a polynomial approximation is linear in the number of edges, making it a suitable approach for large sparse graphs.
- **granularity**: since this metric compares each node's status in the two graphs, this approach benefits from granular information which allows the possibility of identifying the nodes that have undergone the most drastic changes. This is particularly useful in a number of applications where the identification of the area of the graph which changed the most is also of interest (what bacteria radically changed, which neurons adopted a completely different role in the graph, etc.).
- **inclusion of 'mesoscopic' information**: the wavelets allow us to compare neighborhoods of the node at different scales automatically. This enables a less short-sighted representation of the overall graph structure than standard structural distances.
- **inclusion of 'multiscale' information**: the topological signatures that we obtain for each node can be further enriched to contain multiscale information: in [15], a multiscale topological signature associated to scales $\{s_1, \cdots s_j\}$ is defined as the concatenation of the representations : $\chi_a = [\Psi_a^{s_1}, \Psi_a^{s_2}, \cdots \Psi_a^{s_j}]$. In this case, the heat-distance between two graphs is simply computed by replacing $r_a$ in Equation 4.4 by this new value $\chi_a$.This allows for a more robust representation of the topological role assumed by each node. An application of distances based on this multiscale signature on the recipes network is presented in section 6.

These observations also hold for the connectivity-based distances introduced in section 4.1.

4.3. *Application to the microbiome and fMRI study.* Figure 8 shows the results of the analysis of the microbiome study using a heat based distance with $\tau = 1.2$. Interestingly, this distance is one of the few showing a significant link between the years under dependency and the graphs in the fMRI datasets (Friedman-Rafsky for the 5-nn metagraph has a p-value below the 0.05 threshold, Figure8B). However, in the microbiome dataset, this distance is dominated by a clear subject effect (Figures 8B,D and E ). This is further confirmed by the analysis-of-variance test with the subjects as labels described in section 2.2 yields a p-value below $10^{-4}$. We also note that this distance clearly indicates similar dynamics across subjects (Fig. 8C).

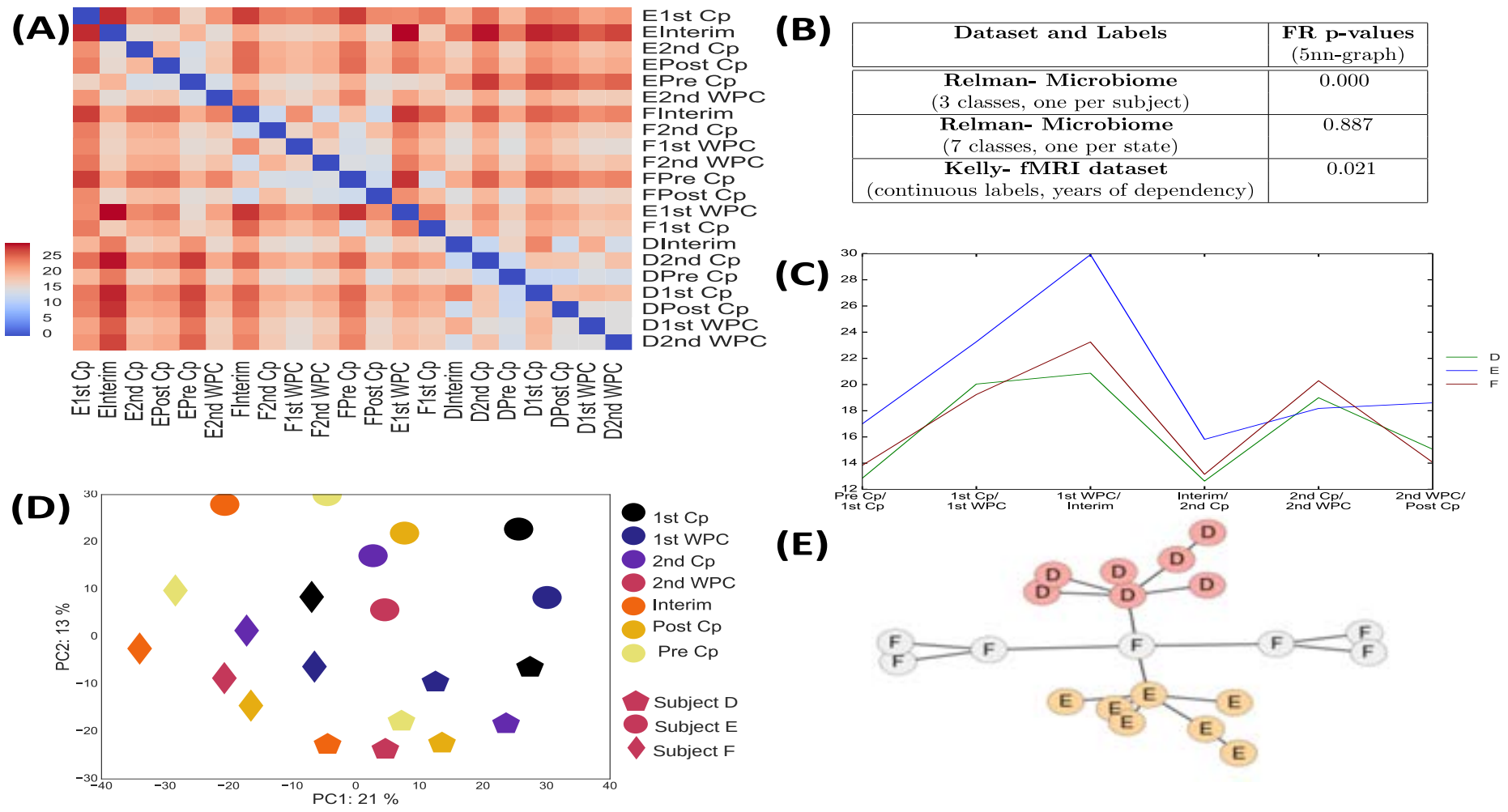

| **Dataset and Labels** | **FR p-values** (5nn-graph) |
|---|---|
| **Relman- Microbiome** (3 classes, one per subject) | 0.000 |
| **Relman- Microbiome** (7 classes, one per state) | 0.887 |
| **Kelly- fMRI dataset** (continuous labels, years of dependency) | 0.021 |

Fig 8: Application of the heat wavelet characteristic distance to the microbiome bacterial graphs, for $\tau = 1.2$. Heatmap of the corresponding dissimilarity **(A)** . P-values of the Freidman-Rafksy test on the 5-nearest-neighbor graphs induced by the heat distance, on each dataset. Plots of the consecutive distances between bacterial graphs **(C)**. MDS projection of the bacterial graphs **(D)** on the first two principal axes. Colors denote treatment phases, and shapes represent different subjects. Minimum Spanning tree induced on the bacterial graphs **(E)**.

## 5. Synthetic Experiments and discussion.

We use several synthetic experiments on toy graphs to enhance our understanding of the different distances' behavior and relative advantages. Synthetic experiments [2] have the benefit of offering a controlled environment for testing the different distances' sensitivity relative to:

- **the graph topology**: we tested the different distances on 5 types of graphs:
  - The Erdös-Rényi model (with $N = 81$ nodes, and a probability of connection $p = 0.1$)
  - a Preferential Attachment (PA) graph on $N = 81$ nodes (with $\alpha = 1.$)
  - a Stochastic Block Model (SBM) graph, with 3 equally sampled communities and connection matrix: $C = \begin{pmatrix} 0.4 & 0.1 & .001 \\ 0.1 & 0.2 & 0.01 \\ 0.001 & 0.01 & 0.5 \end{pmatrix}$

  These sets of network families present different global and local degree densities, and will allow us to assess the impact of the topology on the analysis of network dynamics. The Erdös-Rényi graphs are denser than the preferential attachment graphs, which have an almost star-like structure with a few hubs. The SBM model is somewhere between the two: there are three relatively dense cliques with only a few edges connecting them.
- **the perturbation mechanism**: in our first set of experiments, three initial graphs are generated according to a given topology– ER, preferential attachment, and SBM. We simulate network dynamics as follows: at each time step, for each graph, $\eta$ % of the edges are removed and re-connected elsewhere (at random, following a preferential attachment model). In an attempt to replicate real life situations, we add to this procedure a "background" depletion/thickening process: edges are deleted with probability 0.015 and "formerly absent" edges are added with probability 0.015. How do the distances behave in this simple setting? We expect the curves for the distances depending on topological properties to be stable for denser graphs. In this case, modifications rarely impact the structure of the graph, but structural distances are large, because many edges are being moved. On the other hand, we might observe more instability in the plots for sparser graph structures, where the deletion of a critical edge can have a much stronger impact on the overall connectivity of the graph.

[2]The code for all synthetic and real experiments developed in this review is public and available at: https://github.com/donnate/TrackingNetworksChanges

- **changes in the intensity of the perturbation mechanism**: in our second set of experiments, we want to assess metrics' sensitivity to changes in the dynamical process. At time T=0, we generate on initial graph according to one of our three proposed topologies and simulate a dynamical mechanism as before in which, at each time step, 8.5% of edges are randomly re-wired, and edges are randomly deleted or added with probability 0.015. At $T = 6$, the perturbation mechanism increases its rewiring probability to 34% , and its random deletion/ addition probabilities to 6%. At T=13, the process reverts back to its original characteristics. This induces three distinct time blocks in the time series. The aim here is to see which distances show the existence of a change-point in the graphs' dynamics.

In order to analyze the results, we:

1. quantify different distances' ability to cluster graphs belonging to the same time series: in the first set of experiments, for a low level of noise $\eta$, the distance should recognize graphs belonging to the same time series. To quantify this effect, we use agglomerative clustering on the distance matrix to recover 3 different clusters. We then compute the homogeneity and completeness of these clusters.
2. assess the consistency of the ordering of the graph induced by each distance with the time series: each graph at time $t$ should be closer to its "parent" graph at time $t-1$ and "child" graph at time $t+1$.
3. estimate the ability of each distance to spot changes in the dynamic regime. To this effect, we visualize the heatmaps of the different distances, and compute the ratios $r_1 = \frac{\bar{D}_{12}}{\sqrt{D_{11}D_{22}}}$ and $r_2 = \frac{\bar{D}_{32}}{\sqrt{D_{33}D_{22}}}$, where $\bar{D}_{ij}$ denotes the average "between" (or "within", if $i = j$) distance between graphs in time chunk $i$ and graphs in time chunk $j$.

The legend in each of the subsequent figures indicate the correspondence between curves and distances.

We note that in the SDM case, the eigenvalue-based distances, both the Laplacian and adjacency-based representation yield comparable results.

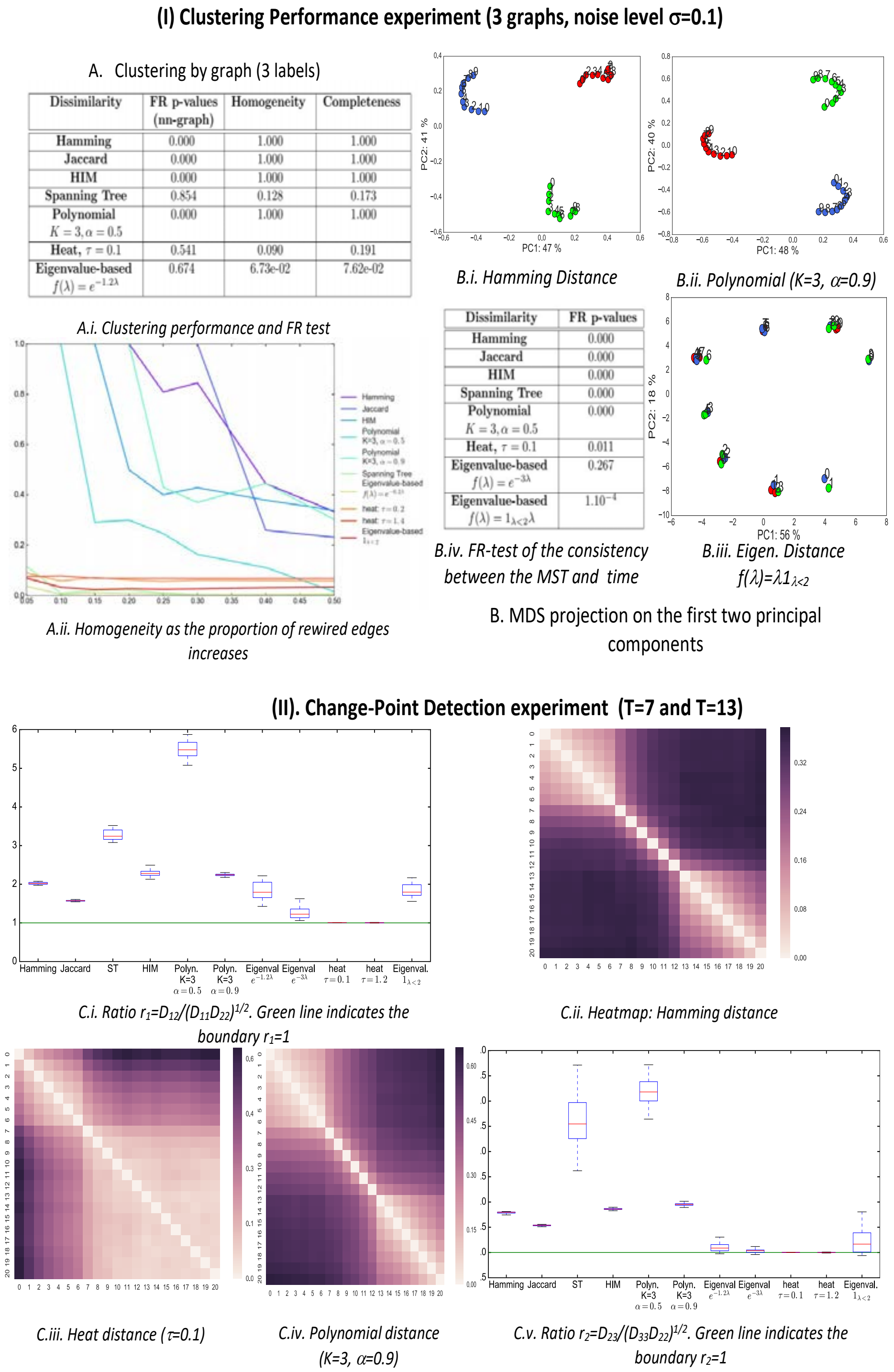

| Dissimilarity | FR p-values (nn-graph) | Homogeneity | Completeness |
|---|---|---|---|
| Hamming | 0.000 | 1.000 | 1.000 |
| Jaccard | 0.000 | 1.000 | 1.000 |
| HIM | 0.000 | 1.000 | 1.000 |
| Spanning Tree | 0.854 | 0.128 | 0.173 |
| Polynomial $K = 3, \alpha = 0.5$ | 0.000 | 1.000 | 1.000 |
| Heat, $\tau = 0.1$ | 0.541 | 0.090 | 0.191 |
| Eigenvalue-based $f(\lambda) = e^{-1.2\lambda}$ | 0.674 | 6.73e-02 | 7.62e-02 |

| Dissimilarity | FR p-values |
|---|---|
| Hamming | 0.000 |
| Jaccard | 0.000 |
| HIM | 0.000 |
| Spanning Tree | 0.000 |
| Polynomial $K = 3, \alpha = 0.5$ | 0.000 |
| Heat, $\tau = 0.1$ | 0.011 |
| Eigenvalue-based $f(\lambda) = e^{-3\lambda}$ | 0.267 |
| Eigenvalue-based $f(\lambda) = 1_{\lambda<2}\lambda$ | $1.10^{-4}$ |

Fig 9: Results for the Stochastic Block Model topology. **Top Row:** Comparison of the smooth dynamics (no change point), with 0.05% edges rewired at each time step. **Bottom Row:** Change point detection experiment.

We summarize our findings as follows:

- **Smooth dynamical process**: We observe that the performance of the different distances is not consistent across topologies. In particular, the dynamic regime associated to the Preferential Attachment graph seems to yield more instability, and none of the distances are able to cluster graphs belonging to the same time series correctly. Nor do they yield an ordering of the graphs consistent with their ordering in the time series. However, in the case of the ER and SBM graphs, all structural distances (Hamming, Jaccard, polynomial), as well as some of the eigenvalue based ones (IM) exhibit perfect recovery of the different clusters and the temporal ordering. This is probably due to the higher graph degree densities. Thus, the addition of an edge perturbs the overall structure less. In this case, it seems that the heat distance as well as the low pass filter emphasize similarities across graphs at a similar "state in time": the MDS projection shows a curve with closely clustered points representing graphs at identical times.
- **Change-point detection problem**: the structural distances exhibit better behavior (as shown by the clear blocks along the diagonal). We note that the ST dissimilarity and the polynomial distance have high $r_1$ and $r_2$ ratios, making them perfect candidates for detecting a change in regime.

Based on the results of this set of experiments as well as the real-life results that were shown in the previous sections, we now conclude with a discussion of the strengths and relative advantages of the different distances.

Overall, structural distances seem particularly fitted for tracking temporal evolutions through time when the nodes' Ids are well defined and hold a particular importance in the network. Indeed, these distances focus on some measure of the volume of edges that change from one graph to the other, and as such, are especially able to recognize graphs belonging to same trajectory. This has proven useful in the microbiome case, where the Jaccard distance was able to recognize a strong "subject effect". The Hamming distance can be further enriched by taking into consideration higher order information and comparing larger neighborhoods with the polynomial distances. However, in real datasets, this distance appears to suffer from the same drawbacks as the Hamming distance: all changes are treated equivalently across nodes, and the distance is too blurred by the numerous changes to correctly capture subtle similarities between graphs. As shown by the synthetic experiments, global spectral or meso-scale distances however seem less fit for the task of recognizing graph trajectories.

However, the real-life experiments have also shown the advantages of including global spectral information in the chosen dissimilarity. This is especially useful in real-life "noisy" setting where the correspondence between nodes from one graph to another is only approximative. In the fMRI dataset typically, the role played by one node in a graph can in fact be assumed by its neighbor the another. In that case, while these graphs will be classified as dissimilar by structural distances, spectral and meso-scale distances are able to recognize the similarity. This is probably what explains the results obtained by the spectral and heat distances on the fMRI dataset, where some interesting associations between the one-nearest neighbor metagraph and the number of years under dependency are detected. The heat-based distance seems to be promising way of achieving "glocality": the scaling factor controls for the propensity of the distance to take into account local information, with low values filtering out "high frequencies" of the signal over the graph and keeping only its low-frequency components. When applied to the microbiome example for instance, heat-distances with lower scaling factors were able to recover the strong subject effect in the data. However, as the scaling factor $\tau$ increases, more global information is taken into account and the distance were then able to recover meaningful treatment stage effects.

## 6. Case study for spatial dynamics: worldwide recipe networks.

In this final section we extend the scope of our analyses from temporal to spatial dynamics through the study of a concrete example: a worldwide recipe network.

In this example, each cuisine is modeled by a graph in which nodes represent ingredients, and edges measure their co-occurrence frequency in various recipes. The motivating intuition behind this graph-based co-occurrence is that cuisines can be better characterized by typical associations of ingredients. For instance, the Japanese cuisine might be characterized by a higher associativity of ingredients such as “rice” and “nori” than Greek cuisine. In our analysis, the graphs were obtained by processing the 57,691 recipes scraped from three different American culinary websites (*allrecipes*, *epicurious*, and *menupan.com*) in [1] as part as a study on food-pairing associations, and counting the co-occurrences of 1,530 different ingredients for 49 different cuisines (Chinese, American, French, etc.)[3]. Each cuisine is then characterized by its own co-occurrence network. The weight on the edge is the frequency of co-occurrence of the two ingredients in a given cuisine. The final graph for a given cuisine thus consists in a collection of disconnected nodes (ingredients that never appear in a single recipe for that cuisine) and a weighted connected component. The construction of these graphs is further discussed in Appendix C.

[3]The data can be downloaded at the following link http://yongyeol.com/pub/

The goal of this analysis is to show which meaningful similarities can be captured by our different distances, and to highlight which is distances are better suited to the comparison of the different cuisine-graphs in this very sparse and unbalanced setting. In this case, natural groupings of cuisines are intuitive, and the results are thus easy to benchmark. Here, we use our inferred pairwise distance matrix between ingredient co-occurrence networks and evaluate our results by plotting both the heat maps of the pairwise distances and constructing "3-nearest-cuisine metagraphs" for each type of distance. In this graph, each node corresponds to a given cuisine $c$. The neighbors of cuisine $c$ correspond to its three-nearest neighbors with respect to a given pairwise similarity matrix. This yields a directed graph of order 3, which we treat here as undirected – hence the degree of each node can be greater than 3 if the node is among the three nearest-neighbors of several cuisines. This provides a way of filtering the information contained in the distance matrix, and quickly visualizing whether the similarities recovered by the distance make intuitive sense.

**Structural distances.**

We begin by analyzing the similarities captured by the Hamming and Jaccard distances. We note that these two structural distances yield very different results. The Jaccard 3-nearest-cuisine summary graph exhibits an interesting tri-cephalic structure (Figure 10b): almost every node in the graph is connected to three main hubs (American, French and Italian). This shows that the Jaccard similarity mostly captures the proportion of shared co-occurrences (as opposed to other network properties). Indeed, here, the American, Italian and French cuisines have the largest connected components (Figure C1c), hence the overlap with the other cuisines' connected components is greater. As such, the Jaccard distance fails to recover more subtle structure in the food network. At the other extreme, the Hamming distance recovers more structure than the Jaccard distances: it manages to recover clusters corresponding to East Asian and East European cuisines. We note here that the similarities are linked with the number of shared ingredients between two cuisine. In particular, the Bangladesh cuisine – whose connected component comprises only 22 ingredients) is uniformly far from the other graphs (Figure 10c). The Hamming distance only reflects the overlap in connected components without accounting for the components relative sizes.

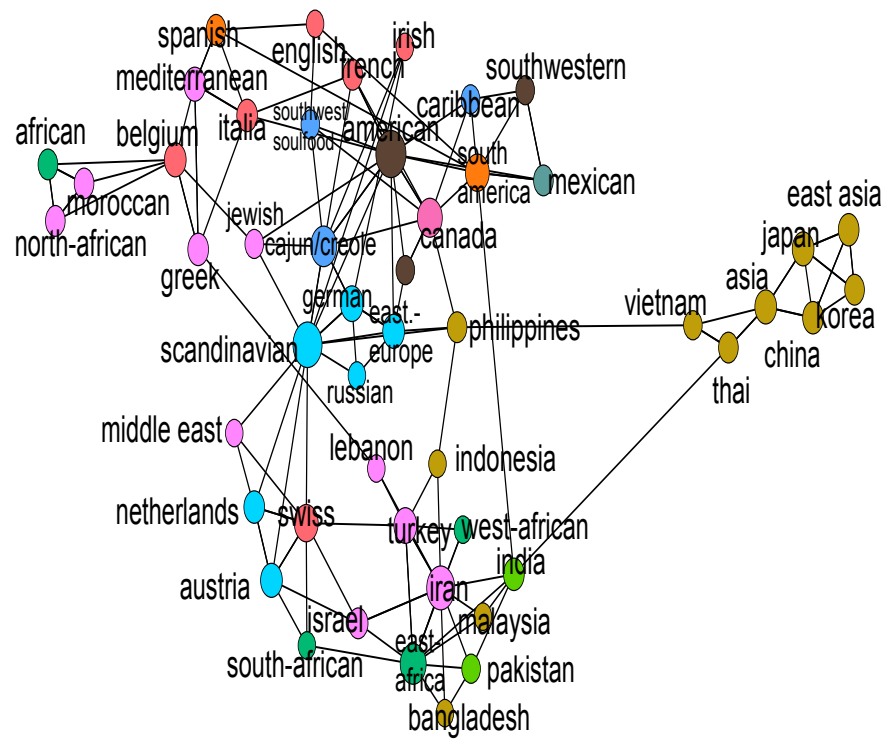

(a) 3-nearest-neighbor Proximity Graph between cuisines (Hamming Distance)

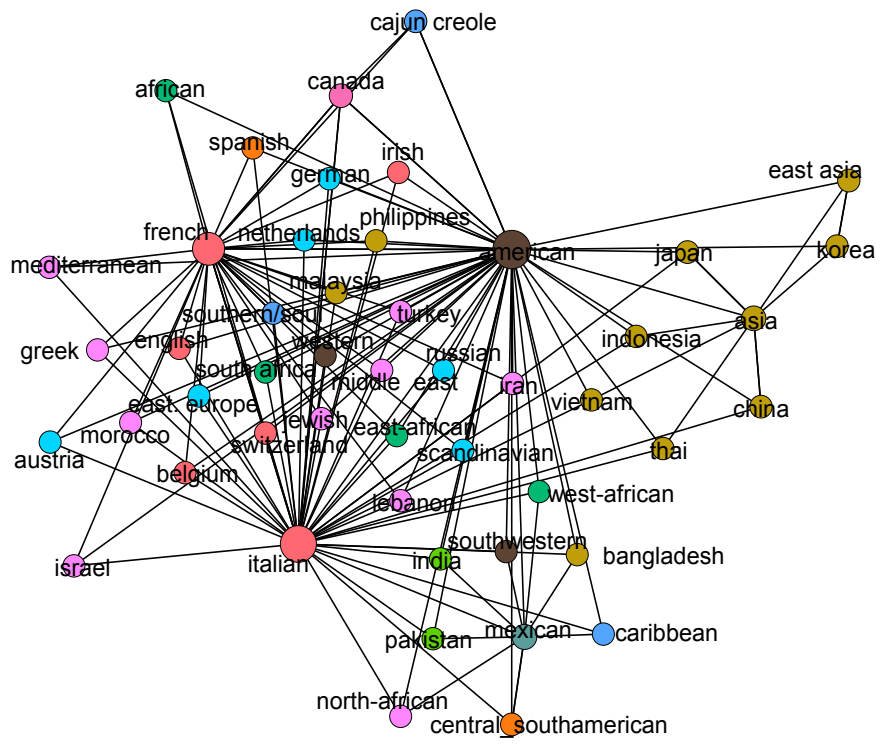

(b) 3-nearest-neighbor Proximity Graph between cuisines (Jaccard distance)

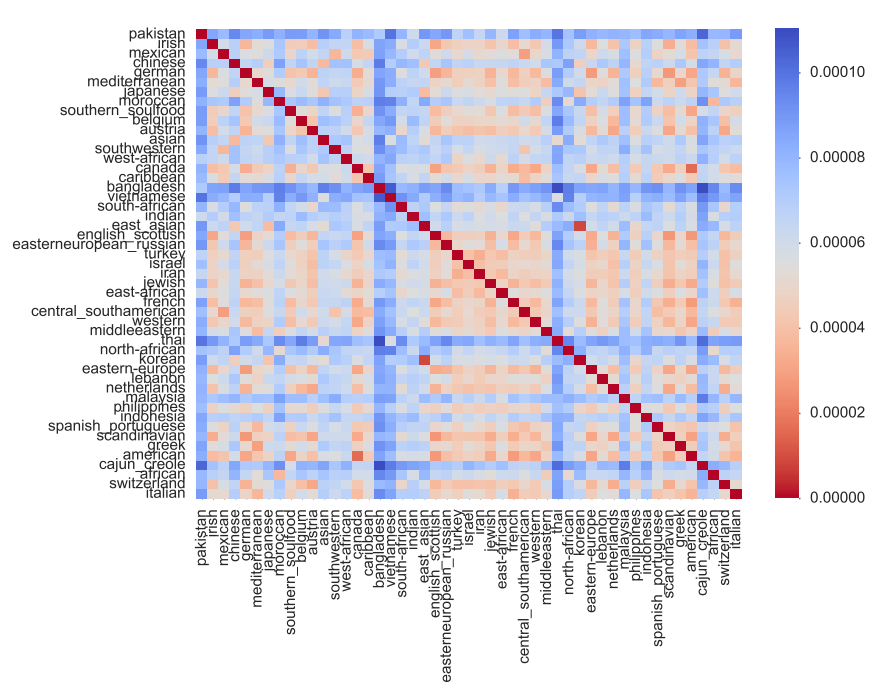

(c) Pairwise distances between cuisines (Hamming Distance)

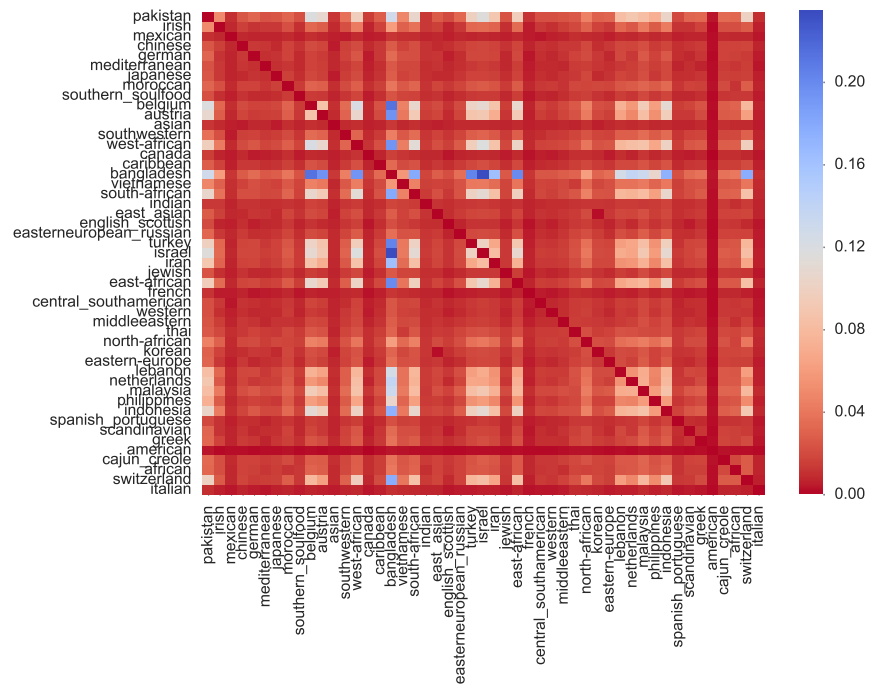

(d) Pairwise distances between cuisines (Jaccard Distance)

Fig 10: Comparison of the pairwise distances and three-nearest cuisine summary graphs (**Left column:** Hamming distance. **Right column:** Jaccard). The three-nearest cuisine summary graph is constructed by representing each co-occurrence network by a node and linking it to its three nearest neighbors according to a given pairwise similarity matrix.

**Spectral distances.** Spectral distances also struggle to recover similarities between cuisines. In this example, the IM and HIM yielded the same 3-nearest-neighbor graphs (Figure 11). We note that this graph consists of two connected components, and does not follow the expected clustering of cuisines: the Mediterranean cuisines for instance (pink nodes in Figure 11) are scattered in each of the two clusters. This might be due to the fact that the IM distance fails in the presence of disconnected graphs, where the eigenvalue 0 has a high order of multiplicity for every graph: in particular, the Bangladesh graph, where 0 has multiplicity 1,508, is very distant from the others. We note that Bangladesh sits unusually far from America, where 0 has order of multiplicity 1,189. For the polynomial distances (Section 3.4), we have taken parameters $\alpha = 0.9$ and $K = 5$ ( a study of $\alpha = 0.5$ and $K = 3$ has achieved the same results). We have also computed a eigenspectrum-based distance (Section 3.1) with $f(x) = e^{-0.9x}$ (using the adjacency matrix of the graph). The polynomial distance seems to recover clusters that are almost consistent with geographical proximities of the different cuisines. However, the lack of structure (no block elements or pronounced groupings) apparent from the heat maps (Figures 12a and 12d) highlights the fact that spectral distances struggle to find definite patterns in this dataset. We thus conclude that a distance based on eigenvalues seems to achieve very unconvincing results for the study of graphs with many disconnected components: in this case, comparing the structure of the graph is insufficient, and we need to include information contained in the nodes' labels.

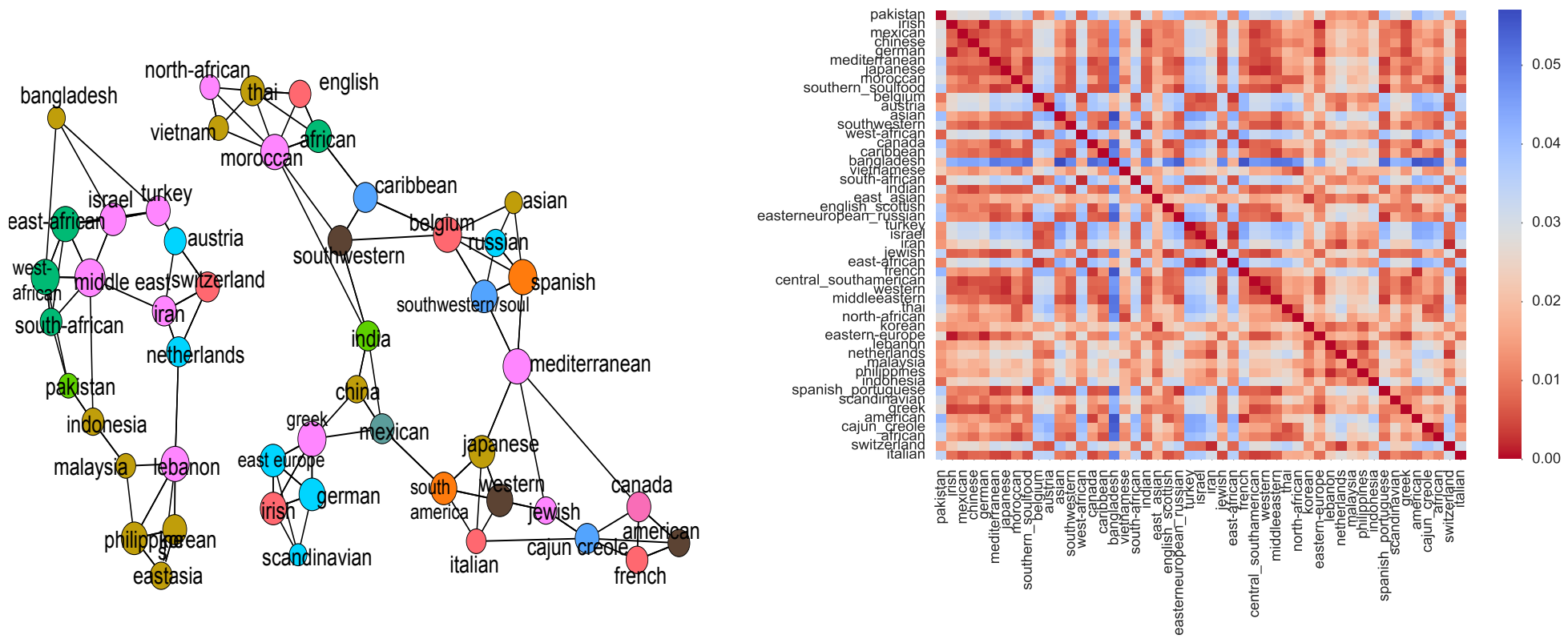

Fig 11: **Ipsen-Mikhailov distance. (right)** 3-nearest-neighbor Proximity Graph between cuisines **(left)** Pairwise distances between cuisines.

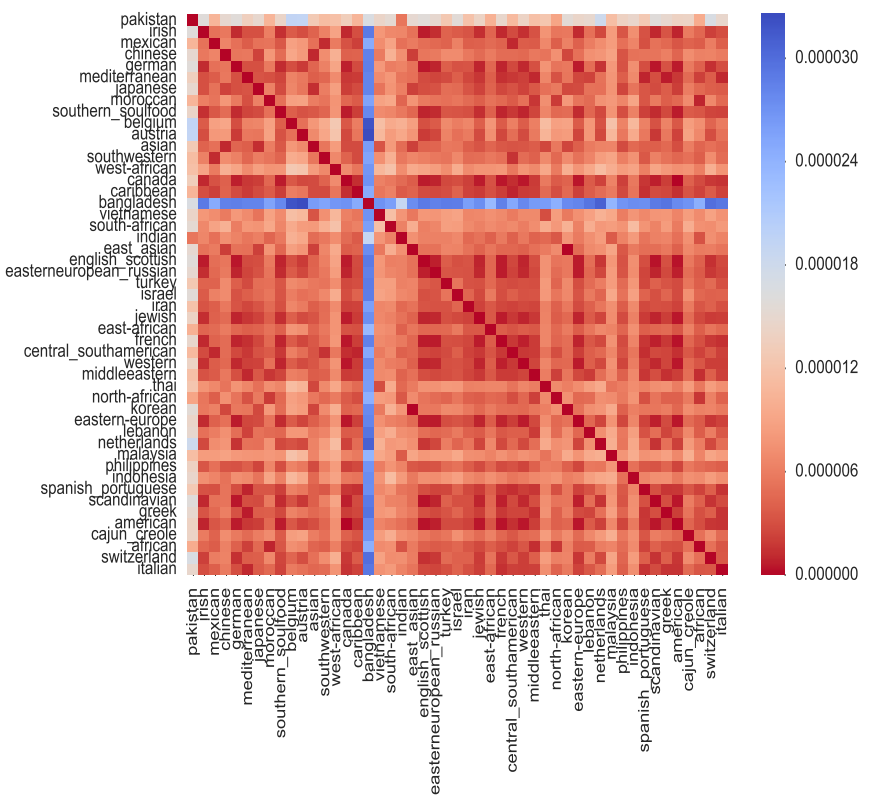

(a) Pairwise distances between cuisines (polynomial distance)

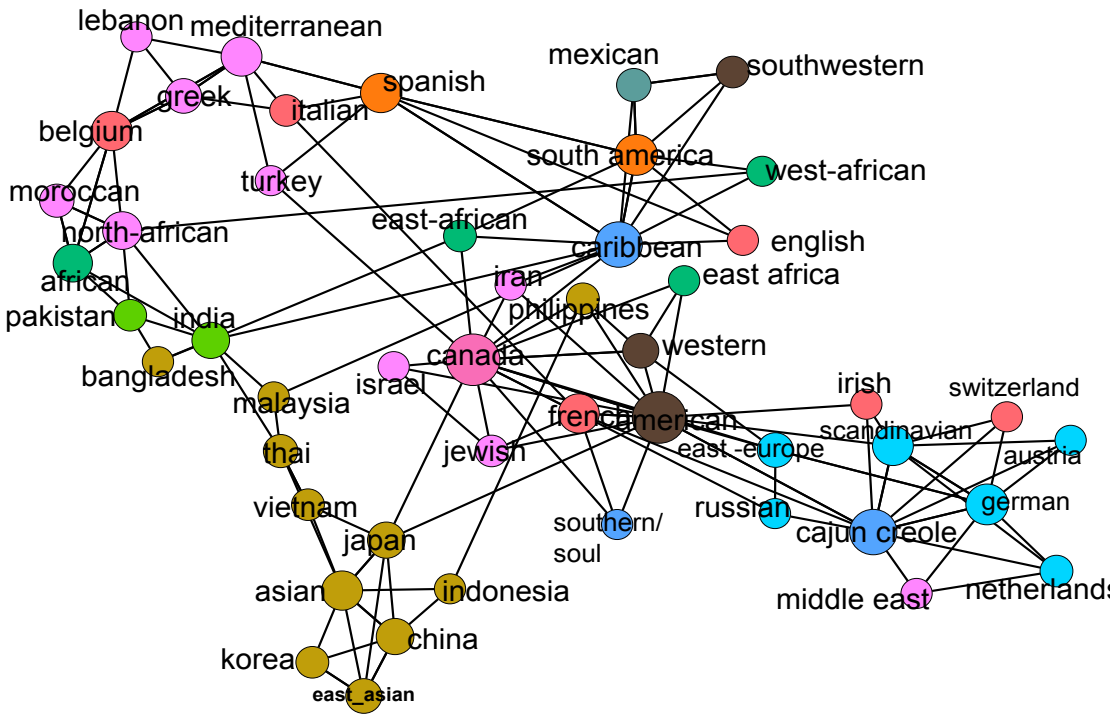

(b) 3-nearest-neighbor Proximity Graph between cuisines (polynomial distance)

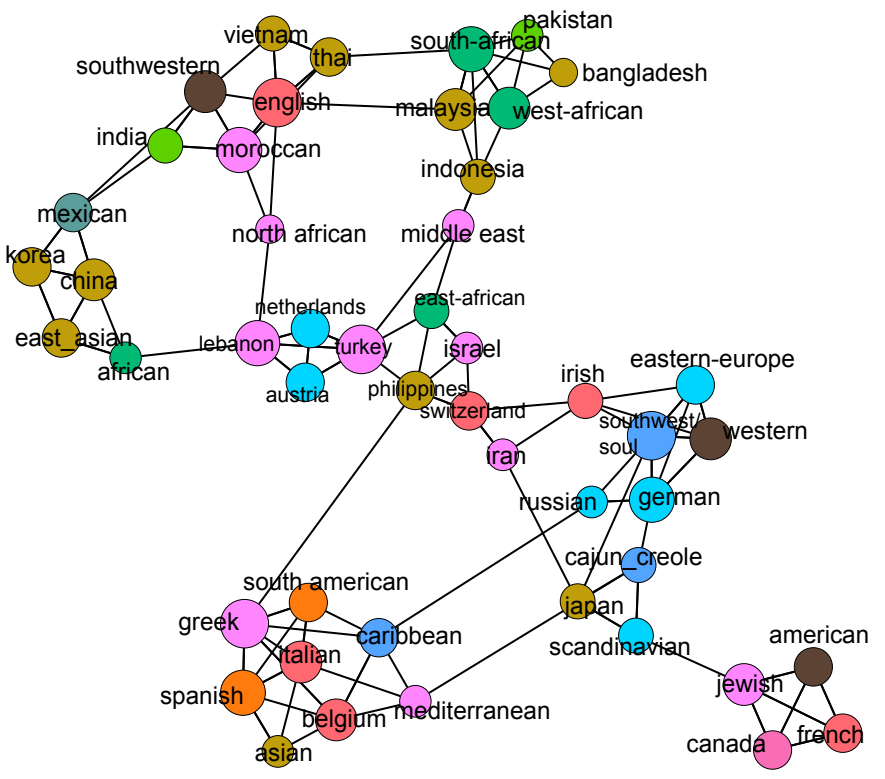

(c) Pairwise distances between cuisines (Eigen Distance)

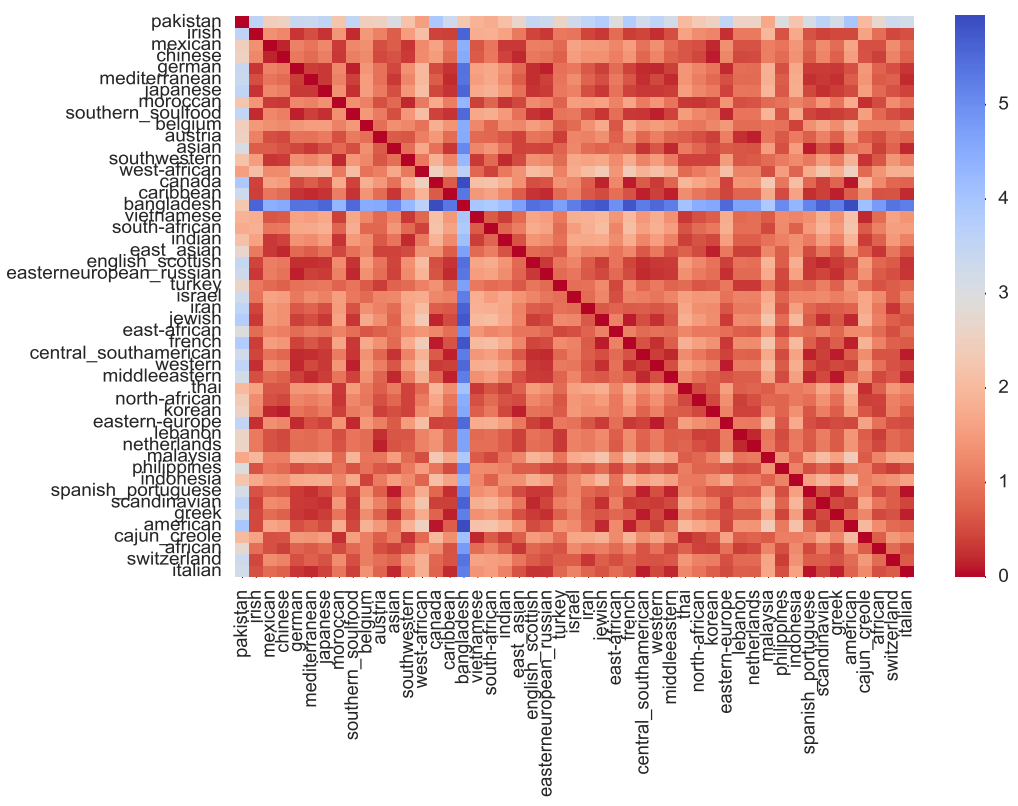

(d) Pairwise distances between cuisines (Eigen Distance)

Fig 12: Pairwise distances between cuisines for various spectral distances. **Top row:** Ipsen-Mikhailov distance. $2^{nd}$ **row:** Polynomial distance (section 3.4), with $\alpha = 0.9$ and $K = 5$. **Bottom row:** Eigenspectrum-based distance (section 3.1) with $f(x) = e^{-0.9x}$.

**Wavelet distances.** In this case, we have computed the heat wavelet signatures for each node according to their multiscale version described in section 4. The scale $s$ was chosen to take values in $\{1, 2, \ldots 29\}$.

Figure 13 shows the 3-nearest cuisine metagraph that this distance yields. We see that the graph that we are able to recover is consistent with geographical proximities would expect. We note for instance the clusters of Scandinavian cuisines and south-western European cuisines, as well as a high proximity of Mediterranean cuisines and Asian cuisines. It is interesting to note that this approach puts Bangladesh cuisine with high centrality. This is due to the limited number of recipes that we have for Bangladesh cuisine yielding more homogeneous and higher edges weights, and thus seemingly closer distance to the other graphs.

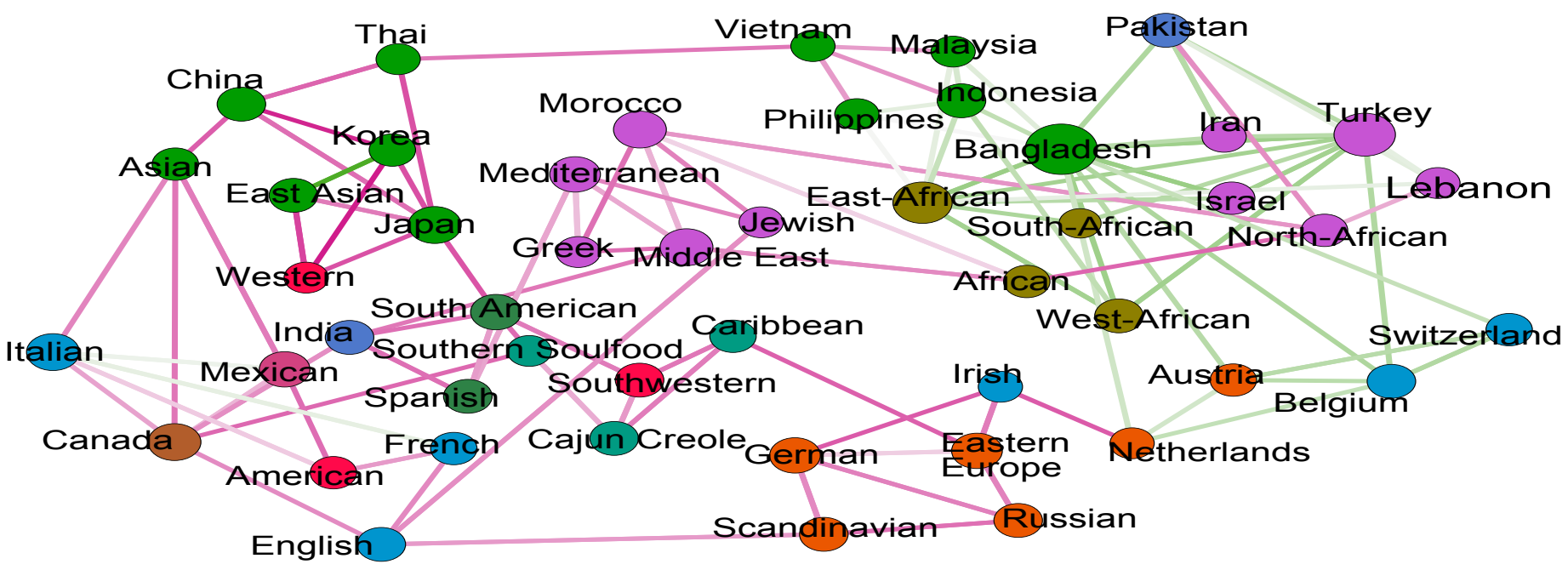

Fig 13: Proximity Graph between cuisines (heat-wavelet based distance)

## 7. Conclusion.

We have given an overview of different metrics and similarity measures for comparing graphs for which we have node labels. Graphs created in real applications rarely have exchangeable nodes and our main focus here has been to supplement the ample literature on permutation invariant graph distances with more refined ones that account for these node identities. Our main focus has been to reflect on the types of changes in dynamic and spatial scenarios that these distances are best suited for. We have provided highlights of these performances on both synthetic and real-data graph analyses.

We have illustrated the use of these metrics for doing statistics on graph-objects in much the same way we did for binary rooted trees in [8]. Finding the right distance for the problem at hand can enable us to further our analyses by constructing the Fréchet mean graph or decomposing the sums of squares of distances between graphs enabling the type of analysis-of-variance type tests that we illustrated in this article.

Distances are useful in assessing many sources of variability in a dataset and as we have shown, can even detect the existence of change-points in dynamics of complex systems such as microbial communities. Pairwise dissimilarity matrices can be used to draw heatmaps (and visualizing the existence – or lack-there-of– of structure in a dataset). Multidimensional scaling embeddings of graphs in Euclidean space allow us to detect latent clusters or gradients.

However, much remains to be done to construct a complete framework for quantifying differences between graphs. For instance, we have not used subgraphs and motif counts that could also be useful in quantifying such similarities as was suggested in [7]. Another possible perspective which we have not covered here focuses on the use of graph kernels [55] to define similarities between graphs.

Finally, we have only taken into account simple node identifiers, whereas incorporating more information about node covariates and edge lengths would enable higher resolution studies and enhanced change-point detection.

**References.**

## Appendices.

### A. *The 2011 Relman microbiome study : a network perspective.*

In this appendix, we present more details about the 2011 Relman microbiome study that serves as one of the guiding thread examples to our discussion.

**The data.** As briefly mentioned in the introduction, this longitudinal study consists of a set of 162 bacterial samples taken from the gut of three distinct subjects (D, E, and F) at different points in time, with roughly the same number of samples (52 to 57) per subject. The subjects were given two courses of antibiotics over a ten month period, yielding seven distinct treatment phases (pre-treatment, first antibiotic course, week after stopping treatment 1, interim, second course of antibiotics, week after stopping treatment 2, and post-treatment phase).

**The Graph Paradigm.** While this dataset has been thoroughly analyzed over the past years [19, 13], we view it through a novel network-based angle. Representation of microbial interactions as a graph has become standard practice[44, 4, 56, 39, 20]. Indeed, bacteria live in symbiosis, feed and proliferate within each body site, yielding potentially complex higher-order interactions. Recent developments in ecology has shown the synergy between bacteria and how it explains various pathologies, such as drug resistant infections [48], inflammatory-bowel disease [27], or diabetes [23]. A deeper understanding of these interactions is thus a necessary step towards more effective translational medicine. In this framework, networks come as a natural tool. They allow to ask a variety of questions such as: are there any significant microbial synergies? Can these be associated to a given pathology? Network structure can also yield insight in the response of the microbial community to perturbations [11, 16].

**Network Inference.** The pre-processing of the data to infer graphs plays a crucial step in the analysis. Microbiome samples are particularly challenging to analyze: generally modeled as zero-inflated negative binomial data, microbiome samples typically exhibit a high number of zero counts. Methods have to take into account the specificity of these data, and a plethora of different methods have been suggested for finding associations between bacteria [18, 5, 38, 57]. We refer the reader to [56] and [39] for a review and comparison of the different methods for inferring networks from co-occurrence data.

In our setting, we do have a small number of samples per treatment phase, and as such, the estimated correlations are inherently noisy. Keeping in mind this noise level, we construct a set of Bacterial "symbiosis" graphs as follows:
• For each subject at a given treatment phase, we define a graph in which each node corresponds to a specific species of bacteria, and edges $\mathcal{E} = \{(i,j)\}$ capture pairwise "affinities" (as measured by the correlation of the abundances through time within each of the different treatment phases) between bacteria $i$ and $j$. Due to the large number of zeros in the data, we use the rank-based correlation metric, Kendall's $\rho$, as a measure of correlation.
• We fix a threshold for each graph: in our case, the threshold that we selected was 0.5, and ensured reasonable sparsity of the network (from 0.02 to 0.3). We emphasize that in this setting, the edges can be "spurious' in the graph, and should not be interpreted as "statistically significant interactions" between bacteria. In fact, these edges are simply a reflection of the co-occurrence between bacteria within different subjects at each different time phase. Figures A1a and A1b show the dynamics in a few cases, highlighting the existence of different synergies for each of the treatment phases.

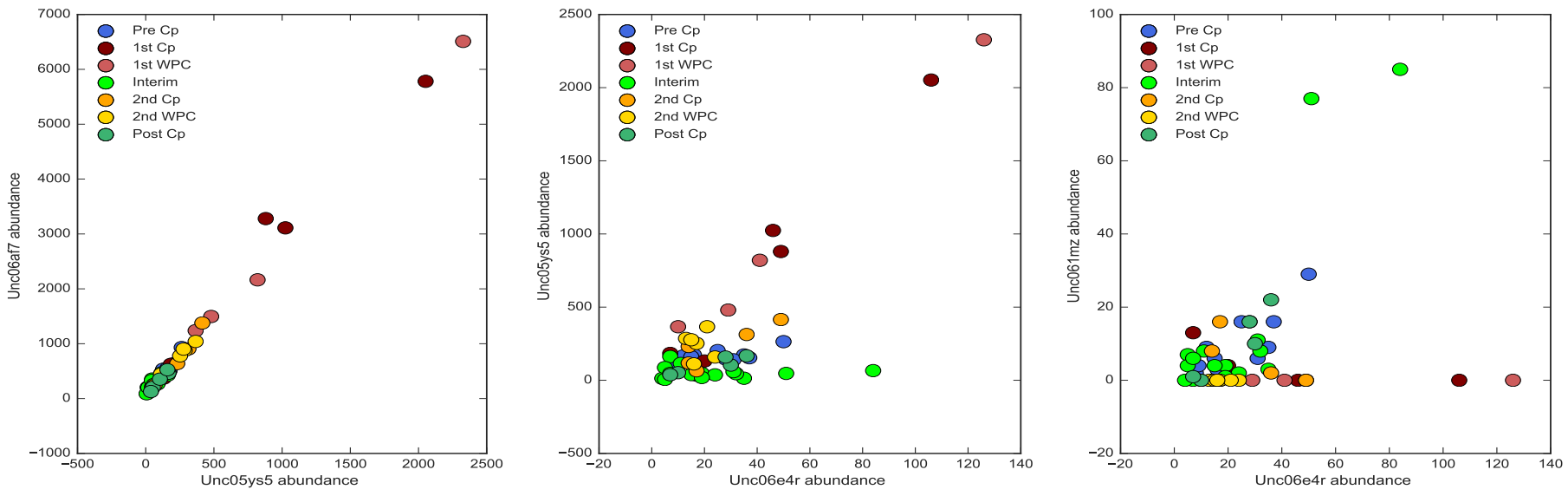

(a) Unc05ys5 vs Unc06af7 (b) Unc06e4r vs Unc05ys5 (c) Unc06e4r vs Unc061mz

Fig A1: **Subject F.** Examples of scatterplots for a few bacteria. Colors denote different treatment phases. The affinities between bacteria vary across treatment phases.

## B. *From fMRI data to brain connectomics.*

In this section, we present more details on the fMRI dataset we used as one of our examples.

The dataset that we have selected consists in the resting-state fMRI of 29 patients under cocaine-dependency. This dataset was gathered by Kelly and co-authors[33] as part of an NIDA-funded grant (R03DA024775) on the impact of cocaine addiction on structural and functional connectivity [4]. This dataset consists in 6-minute resting-state fMRI of patients, pre-processed as per the AFNI and FSL standard pipelines. In particular, the preprocessing included:

- various corrections to account for small head movements and heterogeneity of the fMRI images. These corrections included (as per [33]): slice time correction; 3-D motion correction; temporal despiking; spatial smoothing (FWHM=6mm); mean-based intensity normalization; temporal bandpass filtering (0.009–0.1Hz); linear and quadratic detrending;
- nuisance signal removal (white matter, CSF, global signal, motion parameters) via multiple regression.
- linear registration of functional to structural images (with intermediate registration to a low-resolution image and b0 unwarping)
- nonlinear registration of structural images to the MNI152 template: this allows to align the brain to a common "template" brain.

This yields a total of 116 time-series corresponding to the MNI152 template's nodes, with 140 points each. Consistent with [33], we use these filtered time-series to define a graph based on these series' pairwise Pearson correlation. In order to filter each subject's correlation matrix, we followed an approach akin to [50] and opted for a threshold controlling the graphs' sparsity. The threshold used here is the mean (across subject) of the $97^{th}$ quantile of each patient' correlation matrix (taking only the off-diagonal coefficients). On average, the graphs that we recover have around 3% sparsity, a level in accordance with typical analyses in the field [50] .

This dataset also contains covariates for each subject, with a total 14 variables including gender, age, number of years since first use, smoker, and number of years under dependency. Figure B1 shows the distribution of some of these features.

In their 2011 article [33], the authors use these covariates as evidence of a decreased functional connectivity for patients under cocaine dependence (with respect to healthy control). In this paper, we use our different graph distances to assess whether patients with similar "years of dependency" are more similar.

## C. *The recipes network* .

In this appendix, we present in greater details the recipes dataset analyzed in section 6. The original dataset consists in 57,691 recipes scraped from three different American culinary websites (*allrecipes*, *epicurious*, and *menupan.com*) that were gathered in [1] as part as a study on food pairing associations. Each recipe is labeled by corresponding cuisine (French, American, Greek, etc...). This yields a total of 49 different labels. In this review, we analyzed this dataset from a network perspective by constructing graphs from the co-occurrences of the 1,530 different ingredients in each cuisine. Each of 1,530 ingredients constitutes a node in the graph and each of the 49 cuisine is assigned to a weighted graph. The weight on the edge is the frequency of co-occurrence of the two ingredients for that particular cuisine. We note that in this case, each graph includes a collection of disconnected nodes (ingredients that never appear in a single recipe) and a weighted connected component.

[4]The data is publicly available at the following link http://fcon_1000.projects.nitrc.org/indi/ACPI/html/acpi_nyu_1.html.

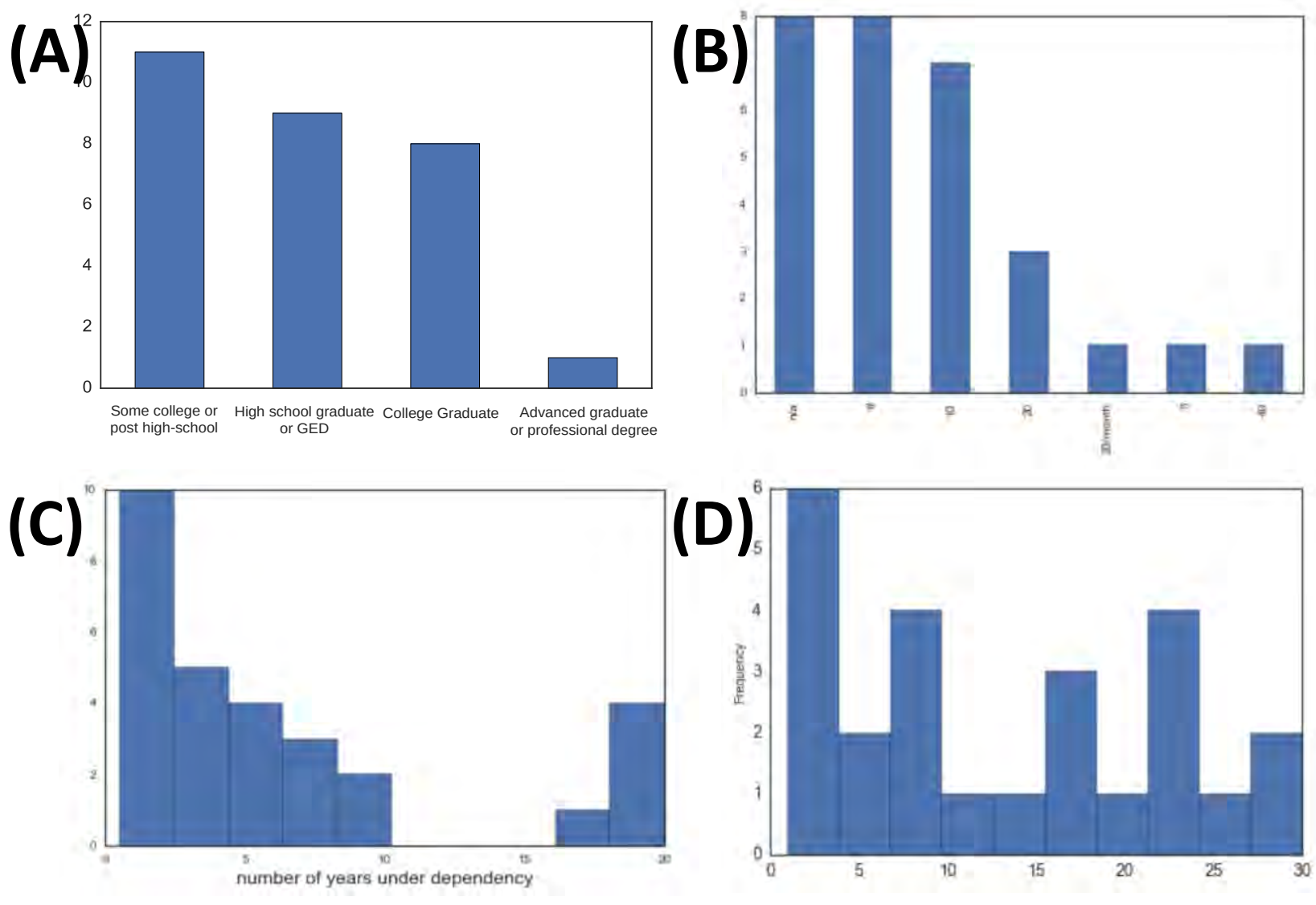

Fig B1: **(A)** Education level. **(B)** Number of cigarettes per day. **(C)** Number of years under dependency. **(D)** Number of years since first use.

Before beginning the analysis of the different graphs, let us quickly highlight the potential challenges of this particular dataset:

- the representation of the different cuisines is highly imbalanced (Figures C1a and C1b). While the American cuisine is extremely well represented (with a little over 40,000 recipes, or 70% of the recipes), conversely, other cuisines are underrepresented: the Bangladesh cuisine, for instance,

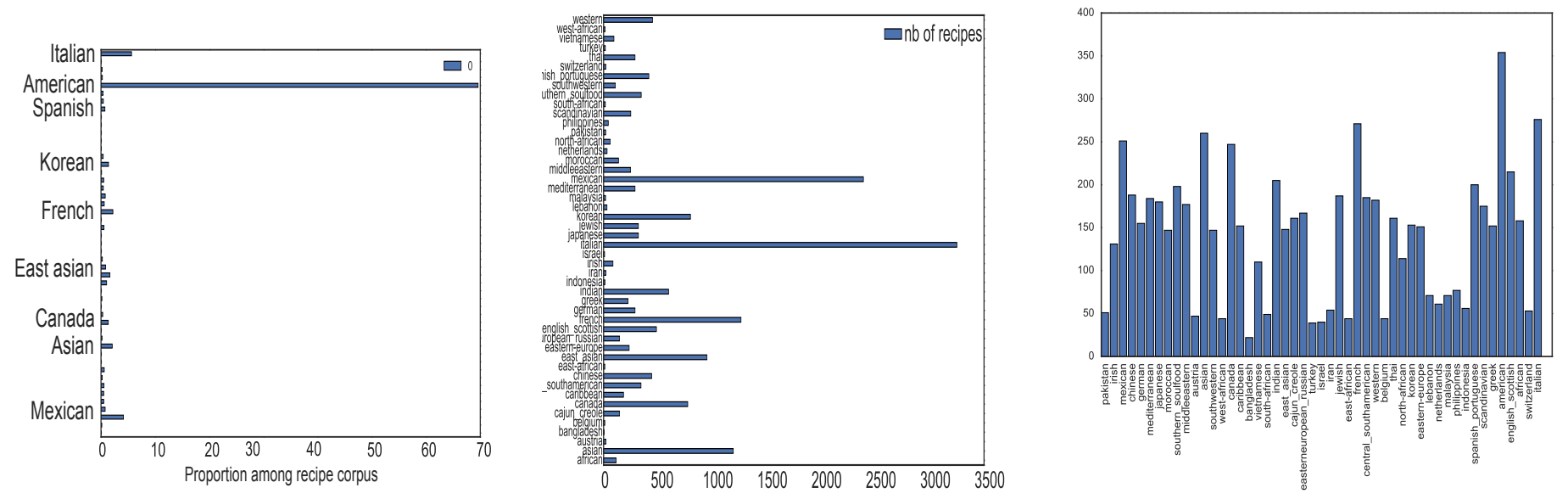

(a) Percentage of recipes from a given cuisine in the corpus. We note the predominance of American cuisine, with more than $2/3^{rd}$ of the corpus.

(b) Number of recipes per cuisine (aside from the American cuisine, for visualization purposes). The minimum is achieved by the Bangladesh cuisine, with only 4 recipes)

(c) Number of ingredients in the connected component. Maximum is achieved by the American cuisine (354 connected ingredients) while minimum is achieved by the Bangladesh cuisine (22 ingredients).

Fig C1: Visualization of the properties of the dataset

only appears 4 times in this corpus. It follows that the number of the ingredients appearing in the connected component of the co-occurrence graphs also varies substantially (Figure C1c).

- Consequently, each cuisine's connected component only accounts for a small fraction of the $1,530$ nodes (Figure C1c). As such, the distance between graphs can be considered very small and the graphs very similar, since more than 80% of the nodes are disconnected from the maximum connected component in both graphs. To account for this imbalance, the distance between cuisine $A$ and $B$ is computed only with respect to the ingredients appearing in either $A$ or $B$'s connected component.

A discussion of the different distances' application to this dataset can be found in the main body of the article. We propose here to discuss an additional benefit of the heat distance over the others–that is, its granularity, which allow to capture in which parts of the graphs the most important changes occurs. Indeed, as detailed in section 8 and in equation 4.3, the heat distance compares in fact the graphs' "topological" signatures, and tracks the variation of the nodes' topological roles from one graph to the other. As such, it is easy to go back to the node level to understand where the variation from one graph to the other is the strongest. We also note that, as described in [15], these structural signatures can be enriched to contain information at multiple scales, yielding a richer "multi-scale" representation of the topological role of each node. Table 1 shows the list of 10 ingredients present in the connected components of two cuisines whose representations have changed the most (from one cuisine to the other). In this case, we have used a multi-scale representation of the signatures ( with scale $\tau \in \{1, \cdots, 29\}$).

| Ingredient comparisons (heat-wavelet based distances) | | |
|---|---|---|
| Cuisine | Neighbor | top changes (char. distance) |
| Middle Eastern | Indian | mustard, dill, bread, thyme, oregano, feta cheese, walnut sesame seed, coconut, olive |
| | Moroccan | chive, nut, red wine, feta cheese, cane molasses, yogurt, rose, oregano, fennel, walnut |
| | Spanish | apricot, lentil, mint, zucchini walnut, pork sausage, feta cheese, sesame seed, lamb, yogurt |
| Chinese | Asian | black bean, oyster, turmeric, cumin, lime juice, nira, coconut, basil, beef broth, lime |
| | Japanese | lemon, oyster, salmon, buckwheat enokidake, tuna,radish, barley, kelp, katsuobushi |
| | Thai | peanut butter, mint, roasted peanut, fenugreek, turmeric, lime juice, cumin, coconut, basil, lime |

Table 1: Identification of the ingredients that change the most from one graph to another

D. *Results for the synthetic experiments.*

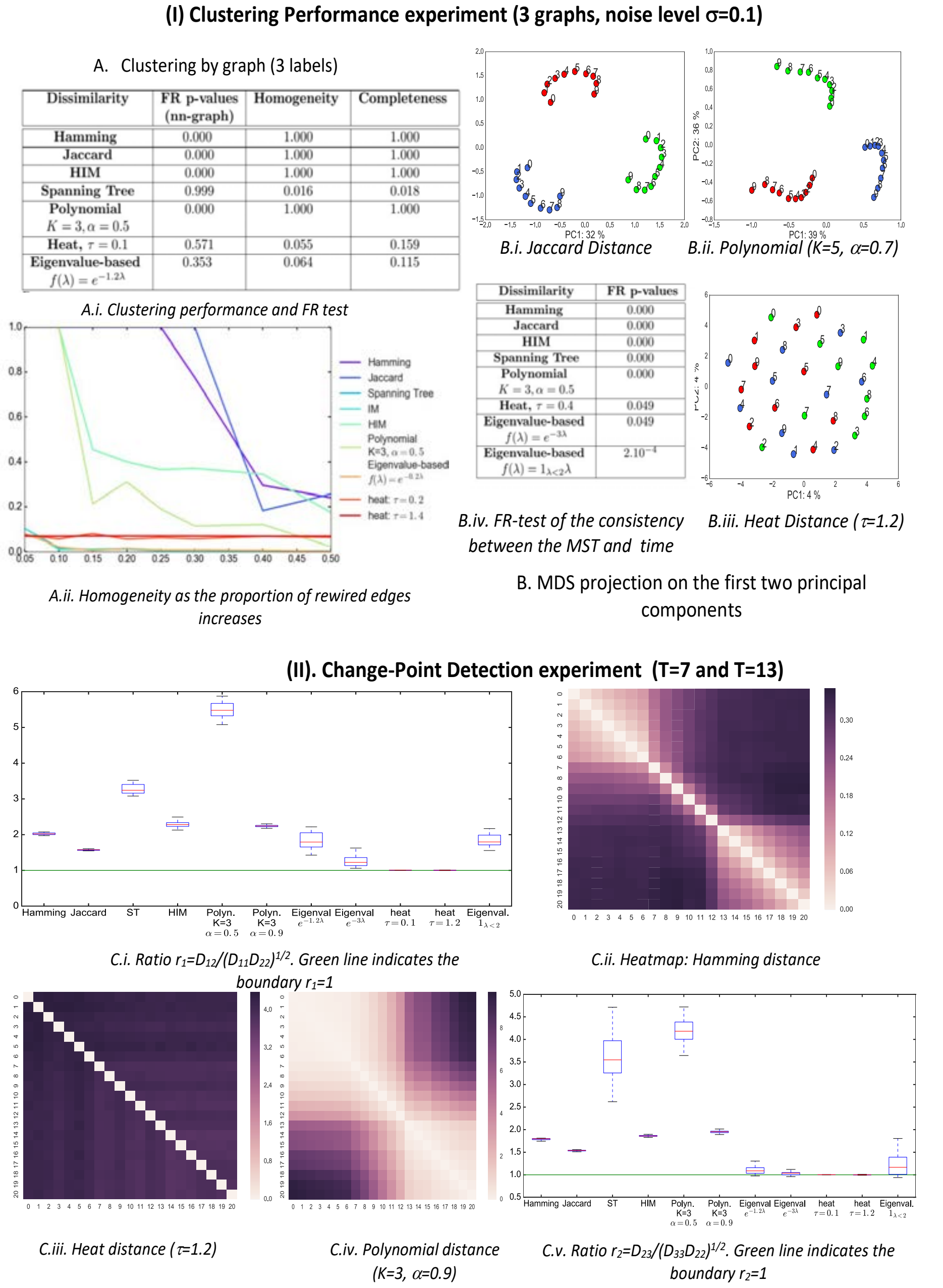

Fig D1: Results for the Erdös-Rényi topology. **Top Row:** Comparison of the smooth dynamics (no change point), with 0.1% edges rewired at each time step. **Bottom Row:** Change point detection experiment.

**(I) Clustering Performance experiment (3 graphs, noise level σ=0.05)**

A. Clustering by graph (3 labels)

| Dissimilarity | FR p-values (3 nn-graph) | Homogeneity | Completeness |
|---|---|---|---|
| Hamming | 1.000 | 00.034 | 0.068 |
| Jaccard | 1.000 | 0.011 | 0.013 |
| HIM | 1.000 | 0.007 | 0.007 |
| Spanning Tree | 1.000 | 0.008 | 0.009 |
| Polynomial $K=3, \alpha=0.5$ | 1.000 | 0.034 | 0.068 |
| Heat, $\tau=0.1$ | 0.996 | 0.008 | 0.009 |
| Heat, $\tau=1.2$ | 0.999 | 0.019 | 0.037 |
| Eigenvalue-based $f(\lambda)=e^{-0.4\lambda}$ | 0.43 | 0.029 | 0.03 |
| Eigenvalue-based $f(l)=1_{\lambda<2}\lambda$ | 0.999 | 0.012 | 0.014 |

*A.i. Clustering performance and FR test*

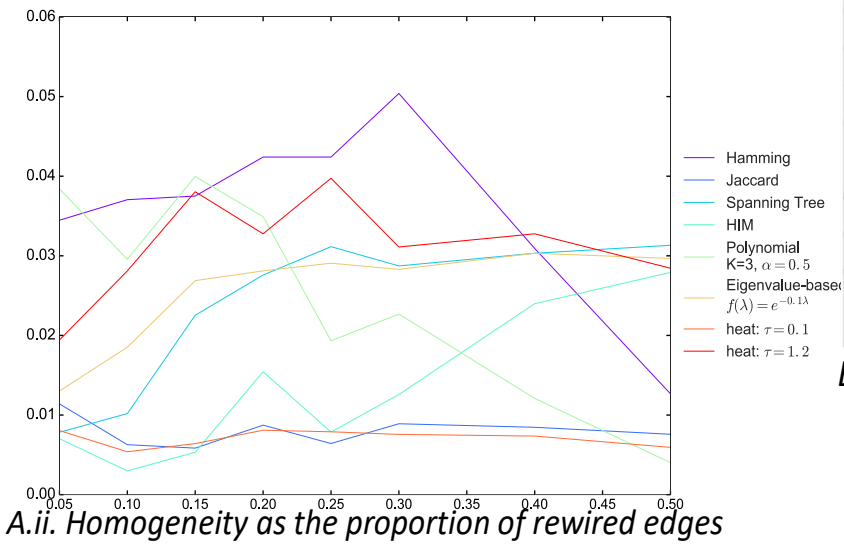

*A.ii. Homogeneity as the proportion of rewired edges increases*

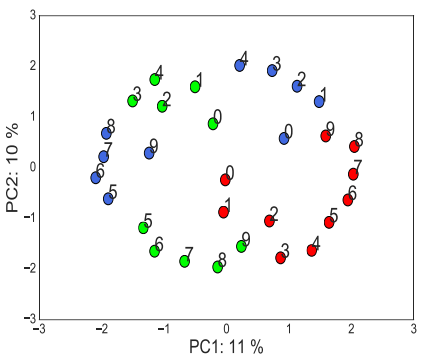

*B.i. Jaccard Distance*

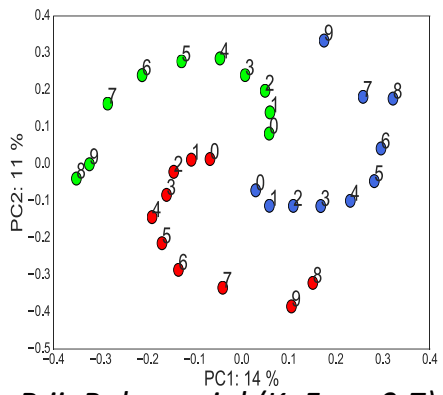

*B.ii. Polynomial (K=5, α=0.7)*

| Dissimilarity | FR p-values nearest-neighbor graph |
|---|---|
| Hamming | 0.999 |
| Jaccard | 0.811 |
| HIM | 0.000 |
| Spanning Tree | 0.000 |
| Polynomial $K=3, \alpha=0.5$ | 1.0 |
| Heat, $\tau=0.4$ | 0.036 |
| Heat, $\tau=1.2$ | 0.981 |
| Eigenvalue-based $f(\lambda)=e^{-0.4\lambda}$ | 0.000 |

*B.iv. FR-test of the consistency between the MST and time*

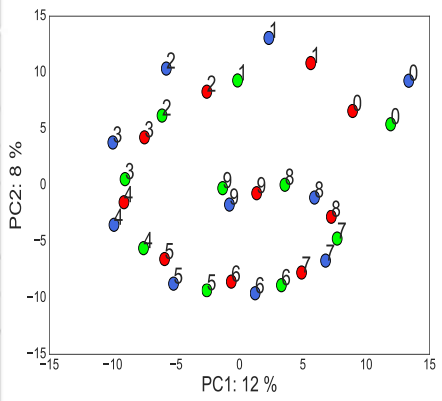

*B.iii. Heat Distance (τ=1.2)*

B. MDS projection on the first two principal components

**(II). Change-Point Detection experiment (T=7 and T=13)**

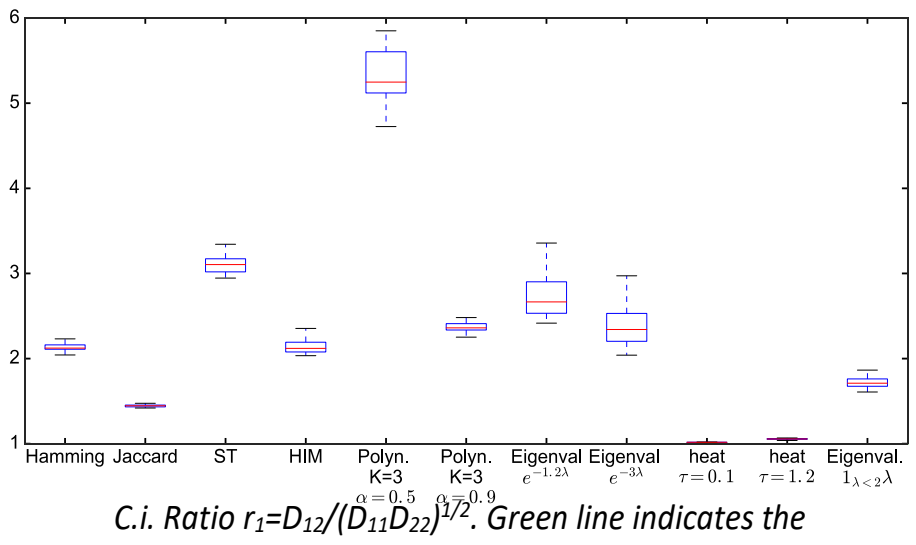

*C.i. Ratio $r_1=D_{12}/(D_{11}D_{22})^{1/2}$. Green line indicates the boundary $r_1=1$*

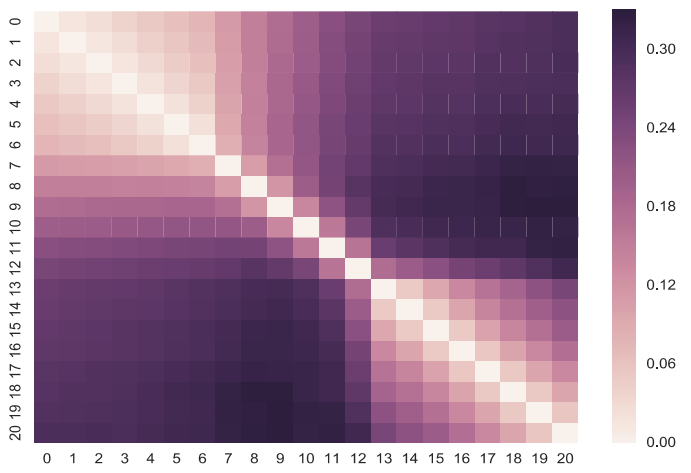

*C.ii. Heatmap: Hamming distance*

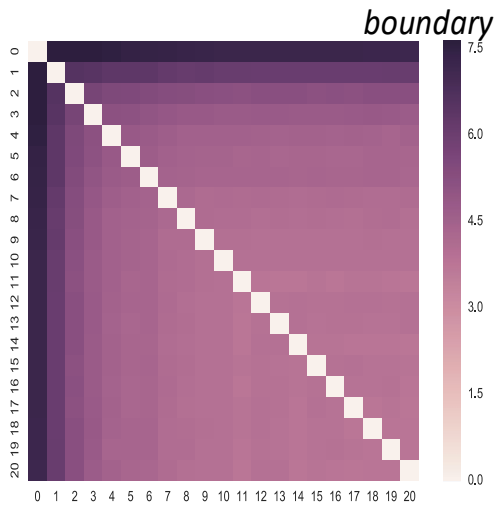

*C.iii. Heat distance (τ=1.2)*

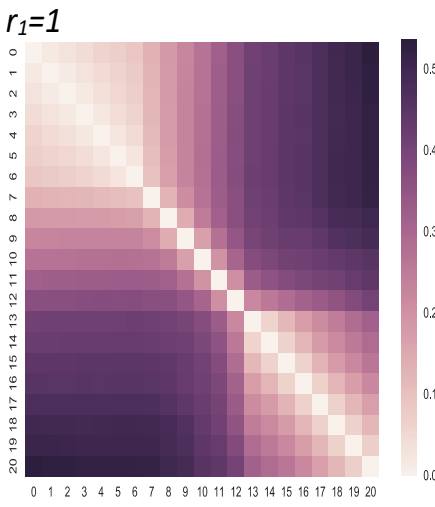

*C.iv. Polynomial distance (K=3, α=0.9)*

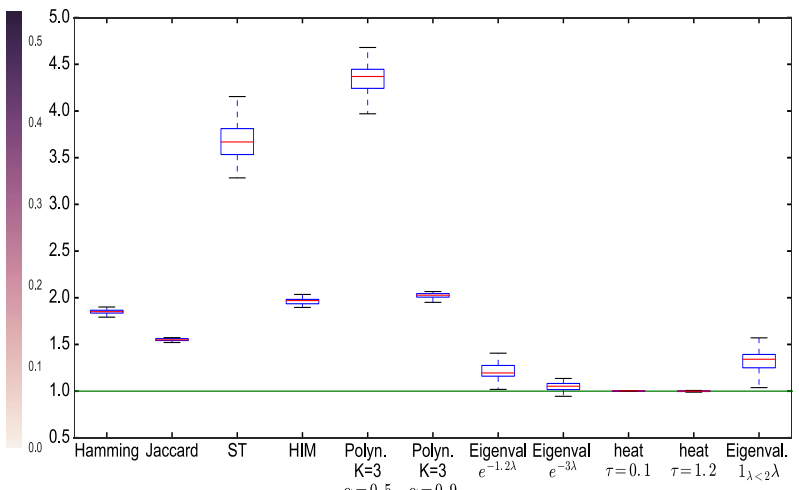

*C.v. Ratio $r_2=D_{23}/(D_{33}D_{22})^{1/2}$. Green line indicates the boundary $r_2=1$*

Fig D2: Results for the Preferential Attachment topology. **Top Row:** Comparison of the smooth dynamics (no change point), with 5% edges rewired at each time step. **Bottom Row:** Change point detection experiment.

E. *Understanding the HIM parameters.*

In this appendix, we give details about the effect of the parameter choices for several of the distances described in this paper.

E.1. *Ipsen-Mikhailov distance.*

Returning to the definition of the Ipsen-Mikhailov distance provided in Eq. 3.3.2. The main drawback of this characterization is that it relies on two parameters, $K$ and $\gamma$, whose values can be approximated but for which the literature (to the best of our knowledge) does not provide any motivation – giving this interesting metric a black-box flavor. To be more explicit and understand how this distance behaves for different types of graphs, let us explicitly determine approximations (in the limit of large $N$) of its parameters.
**Normalization constant $K$.** To begin with, the normalization constant can be written as:

$$\text{(E2)}\quad \begin{aligned} K\int_0^\infty \rho(\omega,\gamma)d\omega = 1 &\iff K\sum_{i=1}^{N-1}\int_0^\infty \frac{1}{\gamma}\frac{1}{1+\left(\frac{\omega-\omega_i}{\gamma}\right)^2}d\omega = 1\\ &\iff K\sum_{i=1}^{N-1}\underbrace{\left[\arctan\left(\frac{\omega-\omega_i}{\gamma}\right)\right]_0^\infty}_{=\pi/2+\arctan(\omega_i/\gamma)} = 1\\ &\iff K = \frac{1}{(N-1)\frac{\pi}{2}+\sum_{i=1}^{N-1}\arctan(\omega_i/\gamma)} \end{aligned}$$

This yields:

- **for the empty graph $\mathcal{E}_n$**, which has eigenvalue 0 with multiplicity $N$:

$$K_{\mathcal{E}_n} = \frac{1}{\frac{(N-1)\pi}{2}}$$

- **for the complete graph $\mathcal{F}_n$**, where $\omega_0 = 0$, and $\omega_1^2 = \cdots =$

$\omega_{N-2}^2 = \omega_{N-1}^2 = N$:

(E3)

$$K_{\mathcal{F}_n} = \frac{1}{(N-1)(\frac{\pi}{2} + \arctan(\sqrt{N}/\gamma))} = \frac{1}{(N-1)(\pi - \arctan(\gamma/\sqrt{N}))}$$
$$= \frac{1}{(N-1)\big(\pi - \gamma/\sqrt{N}[1 - \frac{1}{3}(\gamma^2/N) + o(1/N)]\big)}$$
$$= \frac{1}{(N-1)\pi}[1 + \frac{\gamma}{\pi\sqrt{N}}[1 - \frac{1}{3}\frac{\gamma^2}{N} + o(\frac{1}{N})]) + \frac{\gamma^2}{\pi^2 N} + \frac{\gamma^3}{\pi^3 N^{3/2}} + o(\frac{1}{N^{3/2}})]$$
$$= \frac{1}{(N-1)\pi}[1 + \frac{\gamma}{\pi\sqrt{N}} + o(\frac{1}{\sqrt{N}})]$$

Hence:

(E4)
$$\boxed{K_{\mathcal{F}_n} \approx \frac{1}{2} K_{\mathcal{E}_n}}$$

**The Intuition behind the Scale Parameter.** Through properties of the Lorenz distribution, the scale parameter $\gamma$ is equal to half the interquartile range. It is thus a measure of the "probable measurement error" with respect to the mode $\omega_i$. Here, as proposed by [30], we have chosen $\gamma$ such that the spectral distance between the empty graph and the complete graph is 1: $\epsilon_{\bar{\gamma}}(\mathcal{E}_N, \mathcal{F}_N) = 1$. By definition of the Ipsen-Mikhailov distance, and using Eq.E4 to estimate the different normalizing constants:

(E5)

$$\rho_{\mathcal{F}_N}(\omega,\gamma) - \rho_{\mathcal{E}_N}(\omega,\gamma) = \frac{N-1}{\gamma} K_{F_n}[\frac{1}{(\frac{\omega-\sqrt{N}}{\gamma})^2 + 1} - \frac{2}{(\frac{\omega}{\gamma})^2 + 1}]$$
$$= \frac{N-1}{\gamma} K_{F_n}[\frac{\gamma^2}{N}\frac{1}{1 - 2\frac{\omega}{\sqrt{N}} + \frac{\omega^2}{N} + \frac{\gamma^2}{N}} - \frac{2}{(\frac{\omega}{\gamma})^2 + 1}]$$
$$= \frac{1}{\pi\gamma}\Big(\frac{\gamma^2}{N}[1 - \frac{2\omega}{\sqrt{N}} + o(\frac{1}{N^{1/2}})] - \frac{2}{(\frac{\omega}{\gamma})^2 + 1}\Big) \quad \text{since } (N-1)K_{F_n} \approx \frac{1}{\pi}$$
$$= \frac{1}{\pi\gamma}[-\frac{2}{(\frac{\omega}{\gamma})^2 + 1}] + o(\frac{1}{N})$$

Hence:

$$\begin{aligned}\int_0^\infty [\rho_{\mathcal{F}_N}(\omega,\gamma)-\rho_{\mathcal{E}_N}(\omega,\gamma)]]^2 d\omega &= \int_0^\infty \frac{1}{\pi^2\gamma^2}\frac{4}{((\frac{\omega}{\gamma})^2+1)^2} d\omega + o(\frac{1}{N})\\ &= \frac{4}{\pi^2\gamma}\frac{1}{2}[\frac{\omega\gamma}{\omega^2+\gamma^2}+\arctan(\frac{\omega}{\gamma})]_0^\infty \\ &= \frac{1}{\pi\gamma}\end{aligned} \tag{E6}$$

Hence, here: $\bar{\gamma} \approx \frac{1}{\pi}$

E.2. *Comparison with other spectral distances.*
The advantages Ipsen-Mikhailov distance has two advantages over the other spectral distances: it has a "physical" interpretation as the joint behavior of a system on N strings. Secondly, empirically, it outperforms other spectral distances in capturing regime changes in the dynamics of the network (see the experiments detailed in section 5). However, the running time required by this distance increases dramatically with the number of nodes, making it a less practical tool to work with on large graphs.

The advantages of the Ipsen-Mikhailov distance over the spectral distances previously introduced is linked to its use of the $\ell_2$ norm of the spectral densities rather than raw eigenvalues themselves: the integration of this density makes the distance more robust to small perturbations that have little structural effect, making this distance more appropriate to the analysis of complex systems.

We can assess the impact of the perturbation of a graph $G$, with adjacency matrix $A$. By supposing that the perturbation is small enough that each vibration frequency $\omega_i = \sqrt{\lambda_i}$ of the Laplacian is perturbed by an amount $\epsilon_i$ . Writing $\tilde{\omega}_i = \omega_i + \epsilon_i$, we have:

$$\begin{aligned}\frac{1}{(\frac{\omega-\omega_i-\epsilon}{\gamma})^2+1} &= \frac{1}{(\frac{\omega-\omega_i}{\gamma})^2+1-\frac{2\epsilon(\omega-\omega_i)}{\gamma^2}+\frac{\epsilon^2}{\gamma^2}} = \frac{1}{(\frac{\omega-\omega_i}{\gamma})^2+1-\frac{2\epsilon}{\gamma^2}((\omega-\omega_i)-\frac{\epsilon}{2})}\\ &= \frac{1}{(\frac{\omega-\omega_i}{\gamma})^2+1}\Big[1+\frac{\frac{2\epsilon}{\gamma^2}((\omega-\omega_i)-\frac{\epsilon}{2})}{(\frac{\omega-\omega_i}{\gamma})^2+1}+4\frac{\epsilon^2(\omega-\omega_i)^2}{\gamma^4((\frac{\omega-\omega_i}{\gamma})^2+1)^2}+o(\epsilon^2)\Big]\\ &= \frac{1}{(\frac{\omega-\omega_i}{\gamma})^2+1}\Big[1+\frac{2\epsilon(\omega-\omega_i)}{\gamma^2((\frac{\omega-\omega_i}{\gamma})^2+1)}\\ &\quad -\frac{\epsilon^2}{\gamma^2((\frac{\omega-\omega_i}{\gamma})^2+1)}[1-4\frac{(\omega-\omega_i)^2}{\gamma^2((\frac{\omega-\omega_i}{\gamma})^2+1)}]+o(\epsilon^2)\Big]\end{aligned}$$

Hence, the IM distance between the spectra of graphs $G$ and $\tilde{G}$ becomes:

$$\int_0^\infty [\frac{1}{(\frac{\omega-\omega_i-\epsilon_i}{\gamma})^2+1} - \frac{1}{(\frac{\omega-\omega_i}{\gamma})^2+1}]^2 d\omega$$

$$= \int_0^\infty \frac{4\epsilon^2}{\gamma^2}\frac{1}{[(\frac{\omega-\omega_i}{\gamma})^2+1]^4}\Big[\frac{(\omega-\omega_i)}{\gamma}\Big]^2 d\omega + o(\epsilon^2)$$

$$= \int_{-\omega_i/\gamma}^\infty \frac{\gamma 4\epsilon^2}{\gamma^2}\frac{x^2}{[x^2+1]^4}dx + o(\epsilon^2)$$

$$= \frac{4\epsilon^2}{\gamma} \times \frac{1}{2\times 48}\Big[\frac{(x(-3+8x^2+3x^4)}{(1+x^2)^3} + 3\text{arctan}[x])\Big]_{-\omega_1/\gamma}^\infty + o(\epsilon^2)$$

$$= \frac{\epsilon^2}{24\gamma}\Big[3\pi/2 - (3\text{arctan}(-\omega_i/\gamma) - \frac{(\omega_i/\gamma)(-3+8(\omega_i/\gamma)^2+3(\omega_i/\gamma)^4)}{(1+(\omega_i/\gamma)^2)^3})\Big] + o(\epsilon^2)$$

$$= \frac{\epsilon^2}{16\gamma}\Big[\pi + 2\text{arctan}(\omega_i/\gamma) + \frac{2}{3}\frac{(\omega_i/\gamma)(-3+8(\omega_i/\gamma)^2+3(\omega_i/\gamma)^4)}{(1+(\omega_i/\gamma)^2)^3}\Big] + o(\epsilon^2)$$

$$= \frac{\epsilon^2}{16\gamma}\Big[\pi + 2\text{arctan}(\omega_i/\gamma) + \frac{2(\omega_i/\gamma)}{3}[\frac{-8}{(1+(\omega_i/\gamma)^2)^3} + \frac{2}{(1+(\omega_i/\gamma)^2)^2}$$

$$+ \frac{3}{1+(\omega_i/\gamma)^2}]\Big] + o(\epsilon^2)$$

Summing over all perturbed eigenvalues and taking the square root yields:

$$d_{IM}(A,\tilde{A}) \propto \sqrt{\sum_i \epsilon_i^2} + o(||\epsilon||_2)$$

and the variations are of the order $||\epsilon||_2$. By comparison, provided that the perturbation of each frequency $\omega_i$ is small enough and since $\tilde{\lambda}_i = \lambda_i + 2\epsilon_i\sqrt{\lambda_i} + \epsilon_i^2$, from Eq. 3.1, the standard $\ell_p$-distances are such that:

$$d(A,\tilde{A})^p = \sum_{i=1}^{N-1} |2\sqrt{\lambda_i}f'(\lambda_i)|^p \epsilon_i^p + o(\sum_{i=1}^{N-1}\epsilon_i^p).$$

As such, the distance between $G$ and $\tilde{G}$ puts more emphasis on changes in the eigenspectrum where the product $\sqrt{\lambda_i}f'(\lambda_i)$ is large – which might result in putting too much weight on "noisy" components of the signal, to continue with the signal frequency analogy of section 3.1. Moreover, if we now want to compare the IM distance with the spanning tree distance, we note that the spanning tree (ST) distance is such that:

$$\begin{aligned} d_{ST}(A,\tilde{A}) &= \sum_{i=1}^{N-1} |\log(\lambda_i + \epsilon_i^2 + 2\sqrt{\lambda_i}\epsilon_i) - \log(\lambda_i)| \\ &= \sum_{i=1}^{N-1} |\log(1 + \frac{\epsilon_i^2}{\lambda_i} + \frac{2\epsilon_i}{\sqrt{\lambda_i}})| \approx \sum_{i=1}^{N-1} \frac{\epsilon_i}{\sqrt{\lambda_i}} \qquad (**) \end{aligned}$$

where (**) holds provided that the perturbation remains small compared to the magnitude of the corresponding eigenvalues. However, in the case of sparse graphs, the second smallest eigenvalue (the algebraic connectivity of the graph) is typically very small (and bounded below by $\frac{4}{ND}$ where $D$ is the diameter of the graph). This eigenvalue might thus actually be of the order of the perturbation, thus yielding high variability in the proposed log-ST distance.

The Ipsen-Mikhailov distance can be interpreted as providing an embedding of the distances between graphs in a probabilistic setting, where distributions over eigenvalue densities are compared. This explains its increased robustness to small changes or local perturbations.

390 Serra Mall,
Stanford, CA 94305, USA
E-mail: cdonnat@stanford.edu
susan@stat.stanford.edu